\documentclass[12pt]{article}
\newlength{\abstractwidth}
\usepackage[dvips]{graphicx}
\usepackage{subfigure}

\renewcommand{\thefootnote}{\fnsymbol{footnote}}
\renewcommand{\thanks}[1]{\footnote{#1}} 
 \newcommand{\starttext}{
\setcounter{footnote}{0}
\renewcommand{\thefootnote}{\arabic{footnote}}}
\renewcommand{\theequation}{\thesection.\arabic{equation}}
\newcommand{\be}{\begin{equation}}
\newcommand{\bea}{\begin{eqnarray}}
\newcommand{\eea}{\end{eqnarray}}
\newcommand{\beq}{\begin{equation}}
\newcommand{\ee}{\end{equation}}
\newcommand{\eeq}{\end{equation}}

\newcommand{\<}{\langle}

\renewcommand{\>}{\rangle}
\def\ba{\begin{eqnarray}}
\def\ea{\end{eqnarray}}

\def\14{{1\over4}}
\def\12{{1 \over 2}}
\def\eq{&=&}

\def\h3{h^{3\over 2}}

\def\>{\rangle}
\def\<{\langle}

\def\des{de Sitter Space}

\def\cc{cosmological constant}

\def\0cc{$\Lambda = 0$}
\def\pcc{$\Lambda > 0$}

\def\rd{\dot r}
\def\tt{\tilde \tau}
\def\Om{   \Omega_{D-1}^2   }
\def\om{   \Omega_{D-2}^2   }

\begin{document}
\renewcommand{\theequation}{\thesection.\arabic{equation}}
\begin{titlepage}
\bigskip
\rightline{SU-ITP 04-33} \rightline{hep-th/0408133}

\bigskip\bigskip\bigskip\bigskip

\centerline{\Large \bf {A Framework for the Landscape }}

\bigskip\bigskip
\bigskip\bigskip

\centerline{\it B. Freivogel and L. Susskind  }
\medskip
\centerline{Department of Physics} \centerline{Stanford
University} \centerline{Stanford, CA 94305-4060}
\medskip
\medskip

\bigskip\bigskip
\begin{abstract}
It seems likely that string theory has a landscape of vacua that
includes very many metastable de Sitter spaces. However, as
emphasized by Banks, Dine and Gorbatov, no current framework
exists for examining these metastable vacua in string theory. In
this paper we attempt to correct this situation by introducing an
eternally inflating background in which the entire collection of
accelerating cosmologies is present as intermediate states. The
background is a classical solution which consists of a bubble of
zero cosmological constant inside de Sitter space, separated by a
domain wall. At early and late times the flat space region becomes
infinitely big, so an S-matrix can be defined. Quantum mechanically,
the system can tunnel to an intermediate state which is pure de
Sitter space. We present evidence that a string theory S-matrix
makes sense in this background, and that it contains metastable de
Sitter space as an intermediate state.

\end{abstract}

\end{titlepage}
\starttext \baselineskip=18pt \setcounter{footnote}{0}


 \setcounter{equation}{0}
\section{Introduction}

For string theorists, the importance of making connections with
cosmology is self evident. It would be disappointing to find that
a consistent quantum theory of gravity has nothing to say about
the quantum origin of the universe.  Over the last decade the
lessons from both string theory and black hole physics have
dramatically changed the way we think about space and time
{\it without} seriously changing the way we think about the universe.

However, because of the recent awareness of a large and diverse
Landscape of metastable de Sitter vacua \cite{ls}, the situation may be
changing. A cosmology combining the string-theoretic Landscape
with the ideas of eternal inflation may hold the key to a number
of cosmological puzzles such as the peculiar fine tuning of the
cosmological constant.

In order to give a reasonably rigorous basis to these ideas it is
important to find a framework for studying eternal inflation which
is capable of being adapted to string theory. Thus far, string
theory  requires the existence of an asymptotic boundary on which
some kind of S-matrix data can be defined. The S-matrix
formulation in flat space is familiar. Anti-de Sitter space
provides another space with an asymptotic boundary description. In
both cases the asymptotic boundary is infinite in area and allows
particles to separate and propagate freely.

By contrast, de Sitter space does not allow this kind of
asymptotic description. The space-like asymptotic boundaries of de
Sitter space are problematic. This is particularly so because the
de Sitter vacua of string theory are at best metastable \cite{dsdecay}. This
means that if we follow any time-like path through the geometry we
eventually will find the de Sitter vacuum decaying. The future
time-like boundary is not well described by classical eternal de
Sitter space. In particular, tunneling transitions can occur,
creating expanding bubbles of other vacua. Typically the bubble
nucleation rate is very small and the inflation rate in the
surrounding vacuum is very large. The result is that the space
between bubbles inflates so fast that each bubble remains isolated
beyond the horizon of any other bubble. This is the phenomenon of
\it eternal inflation \rm \cite{linde} (see \cite{eternalinf} for a
review and references).

If the interior of a bubble is a vacuum with a positive vacuum
energy, the space inside the bubble will also inflate, albeit more
slowly than the parent space. The expected result is that the
bubble interiors are themselves eternally inflating regions
 spawning more bubbles.
Following a given time-like curve in this bubble bath, an observer
will see a sequence of vacua which can only end when there is no
longer a positive cosmological constant. This can happen in two
ways. The first is that the observer is swallowed by a region of
negative cosmological constant. This inevitably leads to a big
crunch.  But there is another kind of endpoint with a more
optimistic outlook: a bubble of supersymmetric vacuum with exactly
vanishing cosmological constant can end the sequence. In fact some
observers will exit onto the supersymmetric moduli space while
others  crash into singularities.

Let's suppose a bubble forms in which the cosmological constant is
positive but smaller than in the background. The bubble nucleation
is quantum mechanical and can't be entirely represented
by a classical history, but once the bubble forms it evolves in a
classical way.

\begin{figure}[!!!htb]
\centering \subfigure[]{\includegraphics
[scale=.6]{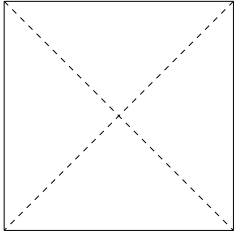} \label{dsPenrose}} 
\hfill 
\subfigure[]{\includegraphics [scale=.6]{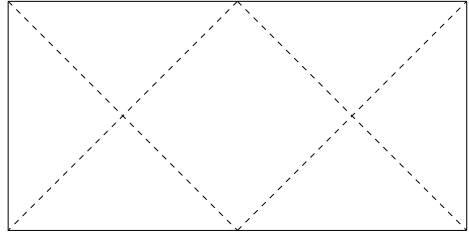}
\label{dsConformal}} \caption{On the left, the Penrose diagram for de
  Sitter space. On the right, the conformal diagram for $1+1$
  dimensional de Sitter space.}
\end{figure}

The process can be illustrated diagrammatically. Let's begin with
the Penrose diagram for  pure de Sitter space in figure
\ref{dsPenrose}. We find it helpful to visualize the $1+1$
dimensional case and to draw the entire space-time as a conformal
diagram as in figure \ref{dsConformal}.

\begin{figure}[!htb]
\centering \subfigure[]{\includegraphics
[scale=.6]{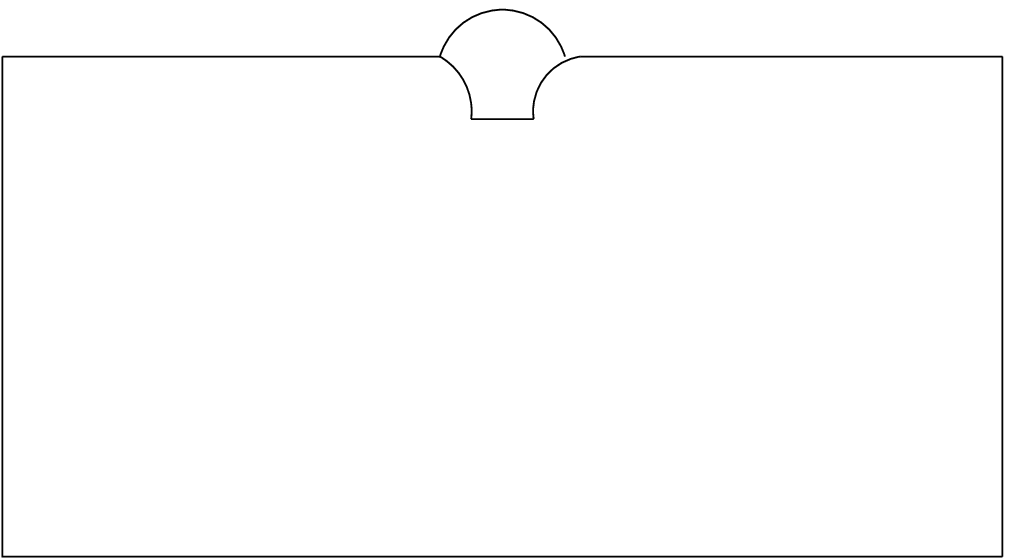} \label{dsbubbleds}} \hfill
\subfigure[]{\includegraphics [scale=.6]{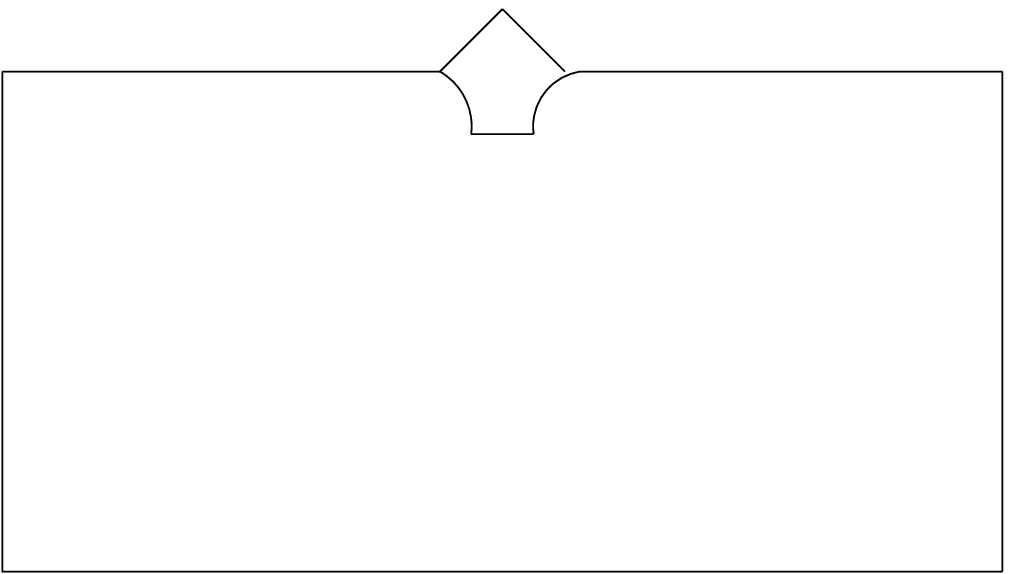}
\label{dsbubbleflat}} \caption{On the left, a bubble of smaller
  cosmological constant forming inside de Sitter space. On the right,
  a bubble of flat space forming inside de Sitter space.}
\end{figure}

 The bubble evolution is shown in figure
\ref{dsbubbleds}. The straight horizontal lower boundary of the
bubble represents the quantum tunneling event. The space-like
upper edge of the bubble is the asymptotic future boundary of the
inflating space in the bubble. If, however, the space in the bubble
has negative cosmological constant then there will be a future
big crunch singularity. Most interesting for us is the case
where the interior of the bubble is a supersymmetric vacuum with
vanishing cosmological constant. Then the diagram looks like
figure \ref{dsbubbleflat}. The future boundary is a portion of the
light-like future infinity of Minkowski space. We will call it a
``hat.'' Only in this case can we describe the future in terms of
well separated asymptotically free particles. That is key to
adapting the method to string theory.

The true asymptotic future will have an infinity of bubbles
forming a fractal set including singularities and hats. The later
the bubble forms, the smaller it will appear on the diagram. The
region near the upper edge of the Penrose diagram is populated
with an infinity of bubbles of all different kinds, separated from
their parent bubble by domain walls. We will refer to the domain
walls as branes. In the case of four space-time dimensions the
branes are 2-branes or membranes.
A curious property of the fractal future is that the coordinate
volume is completely swallowed up by bubbles although the proper
volume of space continues to be dominated by un-decayed de Sitter
space.

At present there is no string theory framework or holographic
framework in which the ideas can be tested. String theory as we
now know it relies very heavily on the existence of an asymptotic
boundary of space-time. The nature of this boundary dictates the
nature of the holographic degrees of freedom as well as the
physical observables. In asymptotically flat space the boundary at
infinity defines the scattering states of string theory. In
anti-de Sitter space the time-like boundary also permits
well-defined boundary data.

By contrast, de Sitter space does not have a clear boundary
description. Most likely eternal de Sitter space is not possible
and in any case there are good reasons to believe that
string-theoretic de Sitter vacua are at best metastable, with a
lifetime shorter than the recurrence time. This means that the
past and future boundaries of de Sitter space have to be replaced
by a quantum fractal of hats and crunches. How exactly string
theory can accommodate this  is far from obvious.

In this paper we suggest a framework for  this purpose. We will
explore the  existence of a background geometry, with two key
properties.

Property one is that there should be asymptotic past and future
boundaries on which the geometry tends to infinite flat space.
This allows us to formulate asymptotic states and an S-matrix.

Property two is that the metastable de Sitter vacua should occur
as intermediate resonant states in the S-matrix.

 \setcounter{equation}{0}
\section{Solutions Interpolating  Between \0cc\ and \pcc\ }

\begin{figure}[!htb]
\center
\includegraphics [scale=.6, clip]
{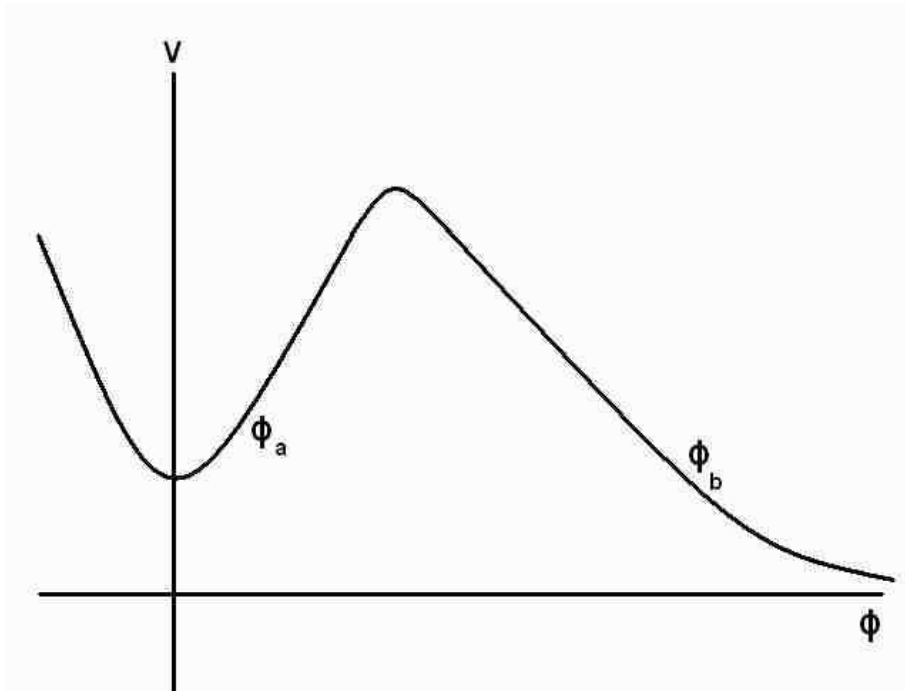} \caption{A potential which has a metastable
  minimum with positive energy and a stable minimum at infinity with
  zero energy.}
\label{eternal-potential.ps}
\end{figure}

Let's begin with a simplified model including gravity and a single
scalar field $\phi$. The scalar potential is assumed to have two
minima, one at $\phi=0$ with a positive value of the vacuum energy
and one with vanishing energy density.  The 
zero energy minimum   can be at finite
$\phi$ or infinite $\phi$. For definiteness we will take the case
of infinite $\phi$, so the potential looks like figure
\ref{eternal-potential.ps}.

If we ignore the effects of gravity then there are spatially
homogeneous solutions of the scalar field equations that closely
resemble the ``S-brane" solutions of open string field theory. The
geometric background is flat space-time and the solutions begin
and end at $\phi=\infty$ where the potential vanishes. The
solutions climb part way up the potential and then roll back down
in a time-symmetric way.

Suppose that the initial kinetic energy of the scalar was close to
the value of the potential at $\phi=0$. In that case one might
wonder if the field can tunnel through to the minimum with
positive energy. If so we would have just the situation that we
are looking for: an asymptotic vacuum with a vanishing \cc \
making a transition to a de Sitter space. If the tunneling rate
is finite the vacuum should eventually tunnel back out of the
potential well and roll back to $\Lambda=0.$

Unfortunately this is not possible. Tunneling cannot happen in a
homogeneous way. If space is infinite, the zero mode of the field
has infinite ``mass." It behaves like a completely classical
coordinate.

A possible way out of this no-go situation is to replace the
minimum at $\phi=0$ by a broad flat region that can support a de
Sitter space, if not forever than at least for a long period.
Eternal inflation  might also be possible. In this case the de
Sitter-like state would decay locally into bubbles that would
indefinitely grow, tending toward the vacuum at $\phi=\infty.$
However if the inflation rate in the undecayed regions is larger
than the decay rate, the regions between bubbles expands faster
than the bubbles. The bubbles remain isolated islands in an
inflating sea. Unfortunately, in this case there is no classical
solution to serve as a background for string theory.

\subsection{ S-Branes}

An interesting possibility is that space-filling unstable D-branes
or brane-anti-brane systems might serve as a starting point for
eternal inflation. We believe this is an interesting avenue to
explore but we will find that nonperturbative methods are
required. As an illustration, consider a system of $N$ unstable
D9-branes in type IIa string theory. The basic requirement for
eternal inflation is that the rate for the open string tachyon to
fall off the top of its potential should be smaller than the
inflation rate at the top of the potential.

Assuming we can ignore corrections, the rate for the tachyon to
fall, call it $\gamma$, is order 1 in string units.
 \be
 \gamma \sim {1\over l_s} \label{fall}
 \ee
 where $l_s$ is the string
length scale.

Let's compare (\ref{fall}) with the inflation rate at the top of the
tachyon potential. The energy density of a D9-brane is $1/g$ in
string units. Thus for a stack of $N$ branes the energy density is
\be \epsilon={N\over gl_s^{10}} \label{epsilon} \ee where $g$ is
the closed string coupling constant. The Hubble expansion rate is
$$
H\sim \sqrt{\Lambda}\sim \sqrt{G \epsilon}.
$$
The  10-dimensional Newton constant $G$ is given by $G=g^2 l_s^8$
giving \be H={1 \over l_s}\sqrt{Ng}. \label{hubble} \ee Thus the
figure of merit, $H/\gamma$, is of order \be {H\over \gamma} \sim
\sqrt{Ng}. \label{merit} \ee

Evidently the ratio of expansion rate to tachyon decay rate is
proportional to the 't Hooft coupling constant for the open string
theory on the branes. It is clear that to push the system into the
regime of eternal inflation the 't Hooft coupling must be at least
of order 1.

Things are not as bad as they could be. The open string theory
must be nonperturbative to exhibit eternal inflation but the
closed string coupling can be arbitrarily small. In other words
eternal inflation may occur in the 't Hooft limit \bea N&\to &
\infty \cr g&\to& 0\cr Ng &\sim& 1 \label{'thooft}. \eea without
any need for closed string quantum corrections.

Although we don't know how to track the system to strong coupling
it is interesting to extrapolate some formulas from weak 't Hooft
coupling to $gN=1$.

First consider the Hawking temperature at the top of the
potential. In de Sitter space the Hawking temperature is of order
the Hubble constant. From (\ref{hubble}) we see that the
temperature at $gN=1$ is
$$
T\sim {1\over l_s}.
$$
In other words the Hawking temperature at the onset of eternal
inflation is the Hagedorn temperature. Perhaps this was to be
expected.

The entropy of a de Sitter horizon is the usual $A/4G$, where $A$
is the eight-dimensional area. The area is
of order
$$
A\sim {1 \over H^8}.
$$
Thus the entropy is \be S \sim {1\over g^6 N^4}={N^2\over(gN)^6 }.
\label{sds} \ee

For $gN=1$ we get the interesting result that the de Sitter
entropy is just the square of the number of branes, \be S\sim N^2.
\ee This is very suggestive since the fields of the open string
theory are in the adjoint representation of $SU(N).$ Perhaps there
is a matrix description of a single causal patch. However it is
not likely that the excited open string states decouple in the 't
Hooft limit. Unlike the case of a system of BPS branes, there is
no reason to expect a decreasing energy scale as $N$ grows. One
reasonable possibility is that the causal patch is approximately
described by the lowest string modes while it is inflating but as
it decays the higher string modes become increasingly important.
Unfortunately we don't have the tools to pursue this further but
it is obviously a worthwhile direction for the future.

\subsection{The Bounce Background \label{bounceback}}

The discussion of the previous section does not in itself help us
find a string theory background with the properties discussed in
section 1, namely, that it permits a description of asymptotic
states composed of freely moving, well-separated particles, and
that it  includes metastable de Sitter vacua as intermediate
states. The homogeneous classical solutions of the coupled
gravity-scalar field equations generally have space-like
singularities either in the past or the future.

Our strategy for finding solutions will be to begin with Euclidean
solutions of the coupled gravity-scalar equations and to continue
them to the Minkowski signature.

The Euclidean solutions we will consider are Euclidean versions of
FRW geometries. They have the form \bea ds^2 \eq c^2dy^2 +a(y)^2
d\Om \cr \phi \eq \phi(y) \label{frw} \eea where $c$ is a constant
and the number of space-time dimensions is $D$.

The variable $y$ runs over a finite range, \be 0\leq y \leq \pi.
\label{range} \ee

The equations of motion are the usual FRW equations apart from
some changes of sign. In 4 dimensions they are: \bea {\dot{a}^2 /
a^2} \eq H^2 = {8\pi G\over 3}\left( \12 \dot{\phi}^2
-V(\phi)\right)  +{1\over a^2}\cr \ddot{\phi} \eq -3H\dot{\phi} +
\partial_{\phi}V(\phi) \label{motion} \eea

 The topology of such solutions is that of the $D$-sphere but the symmetry
 is only that of the $ D-1$ sphere,  $O(D)$. Smoothness at the poles $y=0,\pi$
 requires boundary conditions,
 \bea
a &\to& cy \ \ \ \ \ \ \ \ \ \ (y = 0)\cr a &\to& c(\pi - y) \ \
(y = \pi) \cr
\partial_y \phi &=& 0 \ \ \ \ \ \ \ \ \ \ \ (y=0,\pi)
\label{boundcond}
 \eea

The equations (\ref{motion}),(\ref{boundcond}) are the Coleman-de
Luccia equations \cite{cdl} for the instanton that governs the decay of a
metastable de Sitter space. However at the moment we are
interested in these equations for another purpose. We will
analytically continue the solution to Minkowski signature to give
a classical background.

In order to facilitate the continuation we will write the metric
of the $D-1$ sphere in a form that emphasizes the $U(1)$
rotational symmetry under a particular $U(1)$ in $O(D-1)$. As an
illustration we work with the case $D=4.$ \be d\Omega^2_3 =d\alpha^2
+(\sin^2{\alpha})d\beta^2 +(\sin^2{\alpha}\sin^2{\beta})d\theta^2
\label{3sphere} \ee where $\alpha$ and $\beta$ have  range
$(0,\pi)$ and theta has range $(0,2\pi)$. The $U(1)$ symmetry of
interest is $\theta \to \theta + const$.

The solution (\ref{frw}) will be continued by letting $\theta \to
i t$: \be ds^2=c^2 dy^2 +a(y)^2 \left[ d\alpha^2
+(\sin^2{\alpha})d\beta^2 -(\sin^2{\alpha}\sin^2{\beta})dt^2
 \right].
 \label{mink-continuation}
\ee The angular variable $\beta$ now runs from $0$ to $2\pi$ while
the time-like variable $t$ runs from $-\infty$ to $+\infty$. As in
the Euclidean solution, the field $\phi$ depends only on $y$.

We can rewrite (\ref{mink-continuation}) in the form \be ds^2=c^2
dy^2 +a(y)^2 \left[ d\om -(\sin^2{\alpha}\sin^2{\beta})dt^2
 \right].
 \label{minkcont2}
\ee To get a feel for the geometry we can fix $\alpha$ and $\beta$
and study the geometry in the $y,t$ plane. For example, set
$\alpha=\beta=\pi$. \be ds^2=c^2 dy^2 -a(y)^2 dt^2.
 \label{ytplane}
\ee Now define the conformal coordinate $z$  by 
\bea
c\ dy/a(y) &=& dz  \\
ds^2 &=& a(y)^2[-dt^2 +dz^2] \label{conformalcoord}. \eea

Since $a(y)$ tends to zero linearly at $y=0,\pi$ the range of $y$
is infinite, from $y=-\infty$ to $y=+\infty$. The 2-dimensional
geometry is conformal to the entire $z,t$ plane. Note however that
there is no symmetry under $y\to -y$.

\begin{figure}[!!!htb]
\center
\includegraphics [scale=.7, clip]
{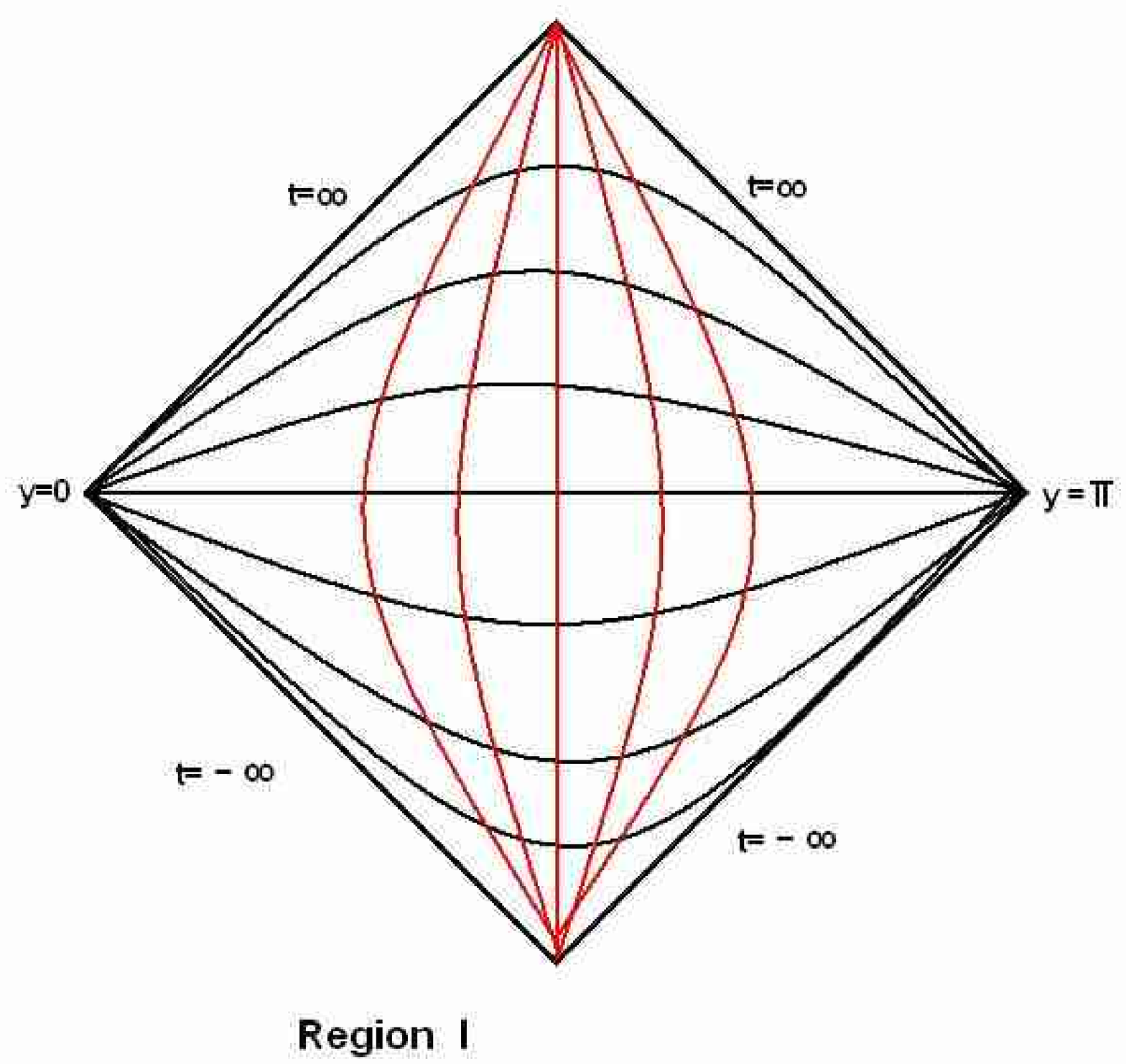} \caption{Wick rotation of the Euclidean geometry
  yields a Lorentzian geometry with this Penrose diagram. The geometry
is not geodesically complete and can be analytically continued to
yield the full Lorentzian solution.}
\label{eternal-region-I}
\end{figure}

We can also draw a Penrose diagram consisting of a diamond as in
figure \ref{eternal-region-I}. In the figure we have superimposed
surfaces of constant $t$. The original rotation symmetry $\theta
\to \theta + c$ is now realized as time translation symmetry.

The geometry however is not geodesically complete. The points
$z=\pm \infty$ are at a finite distance and the light-like
boundaries of the diamond are also not at light-like infinity.
Indeed time-like geodesics cross these light-like surfaces at a
finite proper time.

To complete the geometry let us look at the vicinity of the point
$y=0.$ Letting $\rho=cy$ the metric has the familiar Rindler form,
\be ds^2 = -\rho^2 dt^2 +d\rho^2 \label{rindler}. \ee
As in the Rindler case, the geometry can be continued past
$\rho^2=0$, into the region where $\rho$ becomes a time-like
coordinate-- behind the horizon.  We can continue $\rho^2$ from
positive values (Region I) to negative values in two ways: either
toward the future (into region IV) or the past (region V). In
region I the original $\theta$ shift symmetry became a time-like
translation symmetry but in regions IV and V the symmetry
reverts back to space-like transformations that act on surfaces of
constant (time-like) $\rho$.

The surfaces of constant $\rho$ on the original Euclidean geometry
(\ref{frw})  are obviously $D-1$ spheres. After continuing to the
Minkowski signature they become hyperboloids. In region I the
hyperboloids have a timelike direction and are of a single sheet.
In regions IV and V they are also hyperboloids but they are
space-like. The symmetry ensures that the geometry on each
space-like hyperboloid is homogeneous with uniform negative
curvature. In fact the geometry in these regions has the form of
a conventional Minkowski-signature FRW cosmological geometry with
negatively curved open spatial sections. However in region V the solution
is time reversed relative to region IV.

The equation of motion in region IV is FRW but with a
re-identification of $\rho $ as FRW time and $t$ as one of the
coordinates in the space-like hyperboloids. Recall from
(\ref{boundcond}) that at $\rho=0$ \bea da / d \rho \eq 1 \cr
d\phi/d\rho \eq 0. \label{initial-cond} \eea These now serve as
initial conditions for the FRW evolution.

It is clear that the field in region IV will roll to the point
$\phi=0$ as the FRW geometry evolves. This implies that the FRW
cosmology in region IV is asymptotically de Sitter and terminates
on a space-like infinitely inflated boundary. Moreover region V
is the exact time reverse of region IV.

\begin{figure}[!!htb]
\centering \subfigure[]{\includegraphics
[scale=.6, clip]{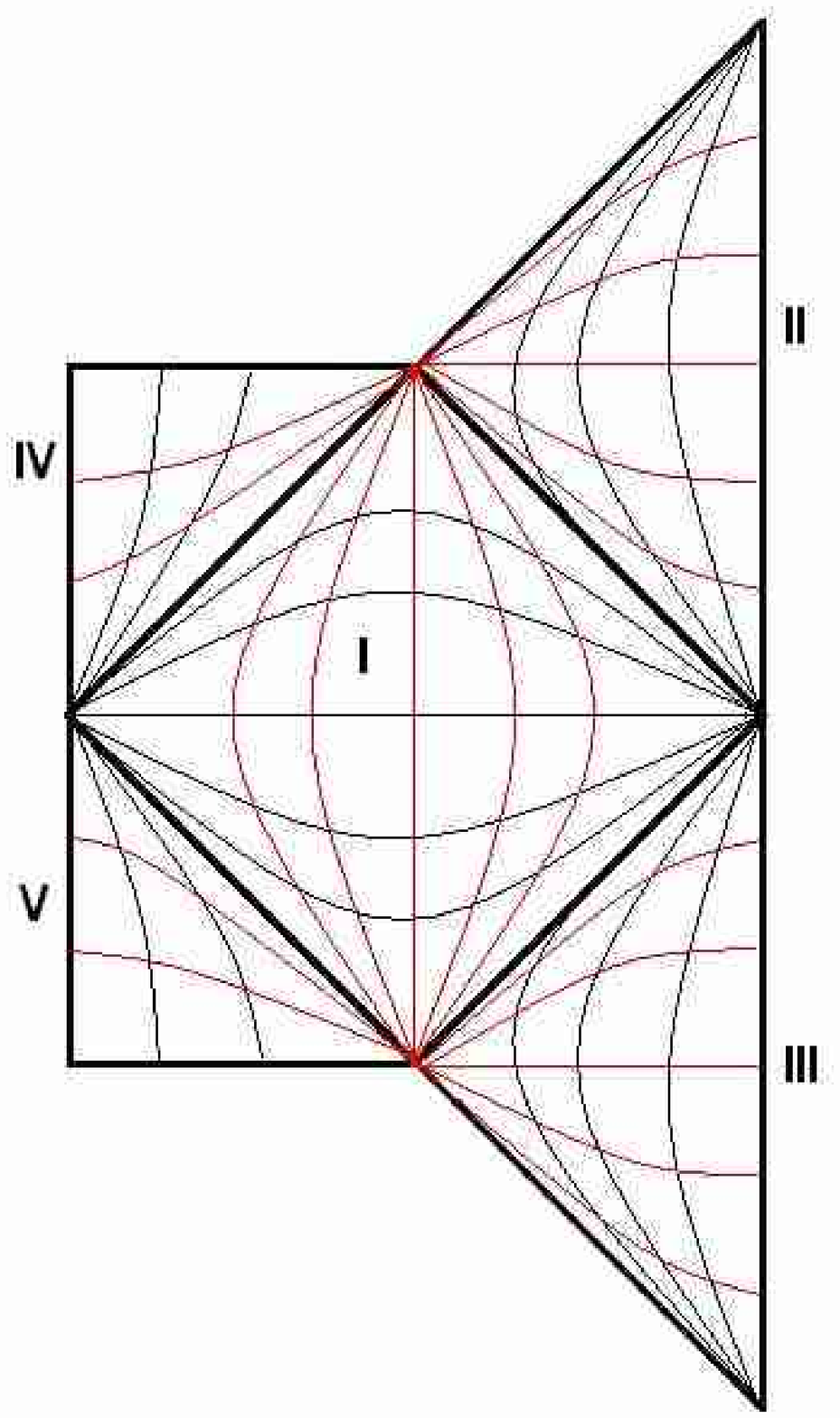} \label{eternal-all-regions}}
\hfill
\subfigure[]
{\includegraphics [scale=.6, clip]
{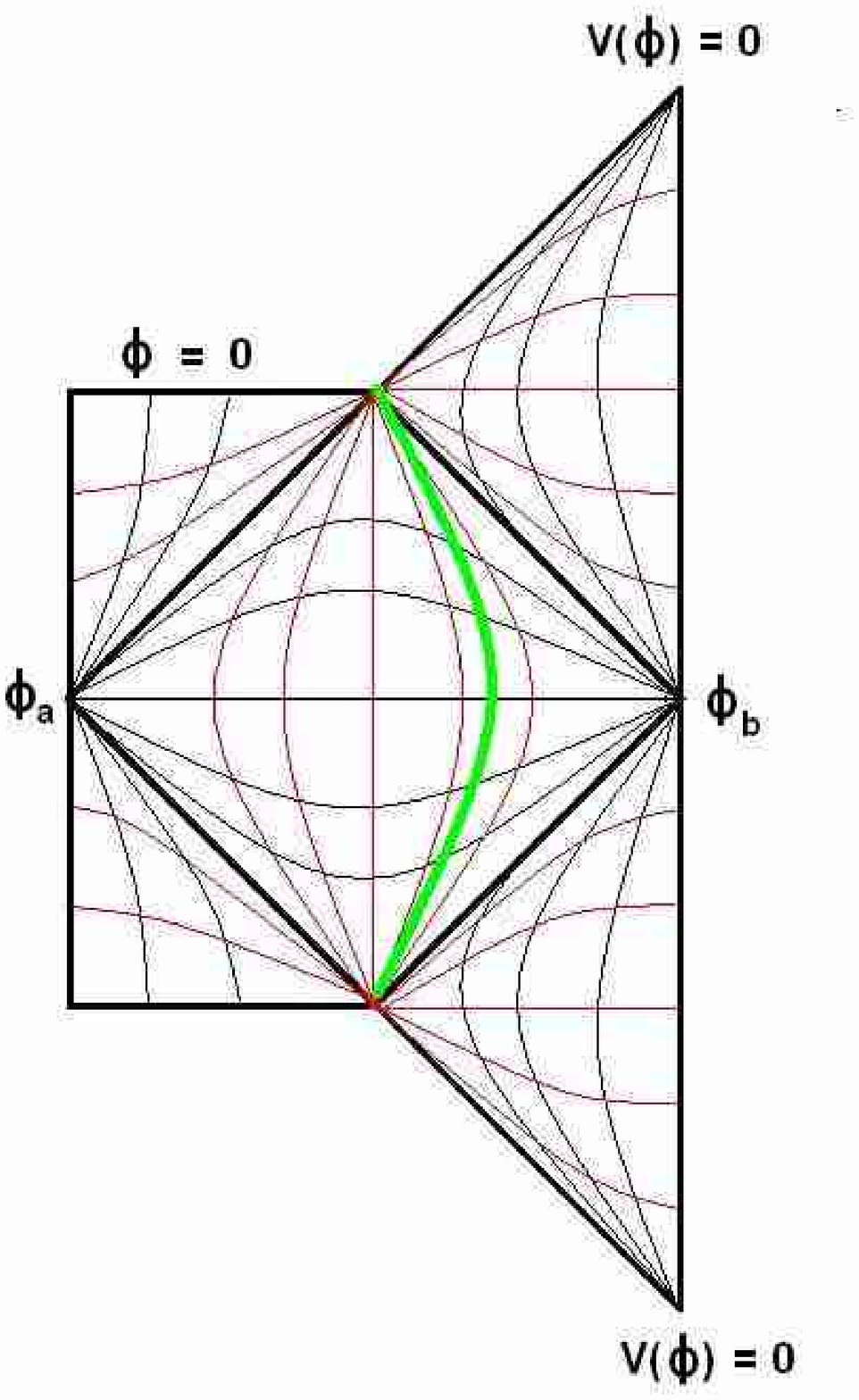} \label{eternal-phi}}
\caption{On the left, the Penrose diagram for the full Lorentzian
  geometry. On the right, the value of the scalar field is shown. At
  the top right of the diagram, the scalar field rolls to its true
  minimum giving \0cc; at the top left the field is in the
  false vacuum and \pcc .}
\label{eternal-phiall}
\end{figure}

Everything we have said about the vicinity of $y=0$ is also true
near $y=\pi$. Continuing past this point is facilitated by
redefining $\rho = c(\pi - y)$. Continuation now produces two new
regions, II (future) and III (past) which evolve as open FRW
geometries--region II in the usual sense and region III in the time
reversed sense. However the initial conditions for regions II and
III are not the same as for regions IV and V. The field starts at
$\phi_b$ on the other side of the barrier. Instead of rolling to
$\phi=0$ it rolls to vanishing cosmological constant. The FRW
geometry will be conventional and end with a time-like and
light-like infinity. In other words the future boundary of the
geometry contains a hat. In figure \ref{eternal-all-regions} we
put all these elements together into a single  conventional
Penrose diagram. Figure \ref{eternal-phi} indicates the values of
the field $\phi$ at various locations.

The interesting thing is that the geometry contains not only a
future hat but also a past inverted hat where well separated
particles can be injected into the system. The past and future
hats will be denoted $h^-$ and $h^+$.

The region of space where the field is near the top of the barrier
defines a domain wall separating the phases where the vacuum
energy is near $V(0)$ or near $0$. Depending of the details of the
potential the domain wall may be thin or thick. In figure
\ref{eternal-phi} the domain wall is the thick green line.

Figure \ref{eternal-phiall} is a conventional Penrose diagram. At
each point there is a local $D-2$ sphere whose radius is a
function of $y$. The radius goes to zero at the two vertical
boundaries of the diagram. We will be particularly interested in
the observer located on the right boundary. Let us consider the
history of such an observer. In the remote past she finds herself
in a contracting FRW geometry with the field $\phi$ rolling up the
potential. The potential reaches its maximum value at $\phi_b$. At
that point the field turns around and the potential starts to
decrease toward $0$. The entire history of this point closely
resembles the behavior of S-branes in open string field theory. In
the  remote past and future FRW regions the field is homogeneous,
but on negatively curved slices and not flat space.

In the frame of the observer, a spherical domain wall  initially
contracts to a minimum radius and then bounces to become an
expanding spherical wall. For this reason we call this solution a
``bounce." As we will demonstrate, solutions of this type provide
backgrounds that satisfy the two criteria mentioned in the
introduction: there are initial and final boundaries where
incoming and outgoing free particles can propagate, and de Sitter
space is  a intermediate resonance in scattering amplitudes.

\begin{figure}[!!!htb]
\center
\includegraphics [scale=.7]
{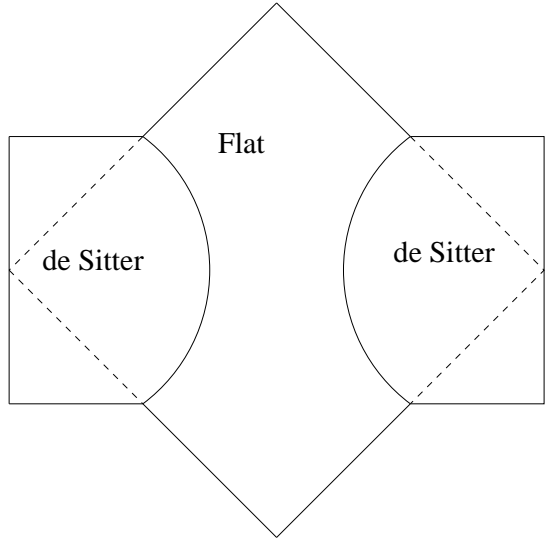} 
\caption{The thin wall limit of the geometry consists of flat space
  separated by a domain wall from de Sitter space. The true Penrose
  diagram is half of the figure. The diamond shaped region is
the causal patch of an observer at
  the center. }
\label{bubbleConfHor}
\end{figure}

The bounce geometry is particularly simple in the thin wall
approximation. The thin wall geometry was discussed for solutions including
the bounce background in \cite{eva}. Similar solutions were discussed in \cite{Giddings}.
The thin wall limit is applicable when the vacuum energy at
$\phi =0$ is very small.  In that case a sharp domain wall
separates the geometry into two domains. For $y< y_{dw}$ the field
is constant and equal to $0$. The space in this region is exactly
de Sitter space. For $y>y_{dw}$ the field is at the minimum where
the vacuum energy vanishes. The space is flat in this domain. The
de Sitter domain includes part of region I and all of regions IV
and V. The flat domain includes the remaining part of region I
as well as regions II and III. This is illustrated in figure
\ref{bubbleConfHor}.

\begin{figure}[!htb]

\centering \subfigure[]
{\includegraphics [scale=1]{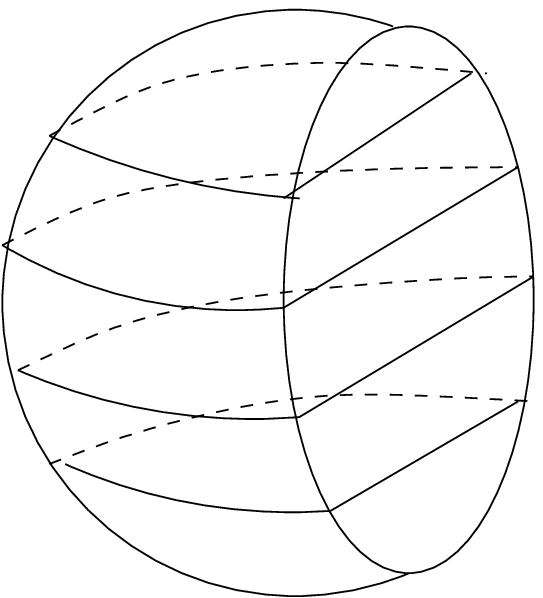}
\label{euclidean}} \hfill
\subfigure[]{\includegraphics
[scale=.6]{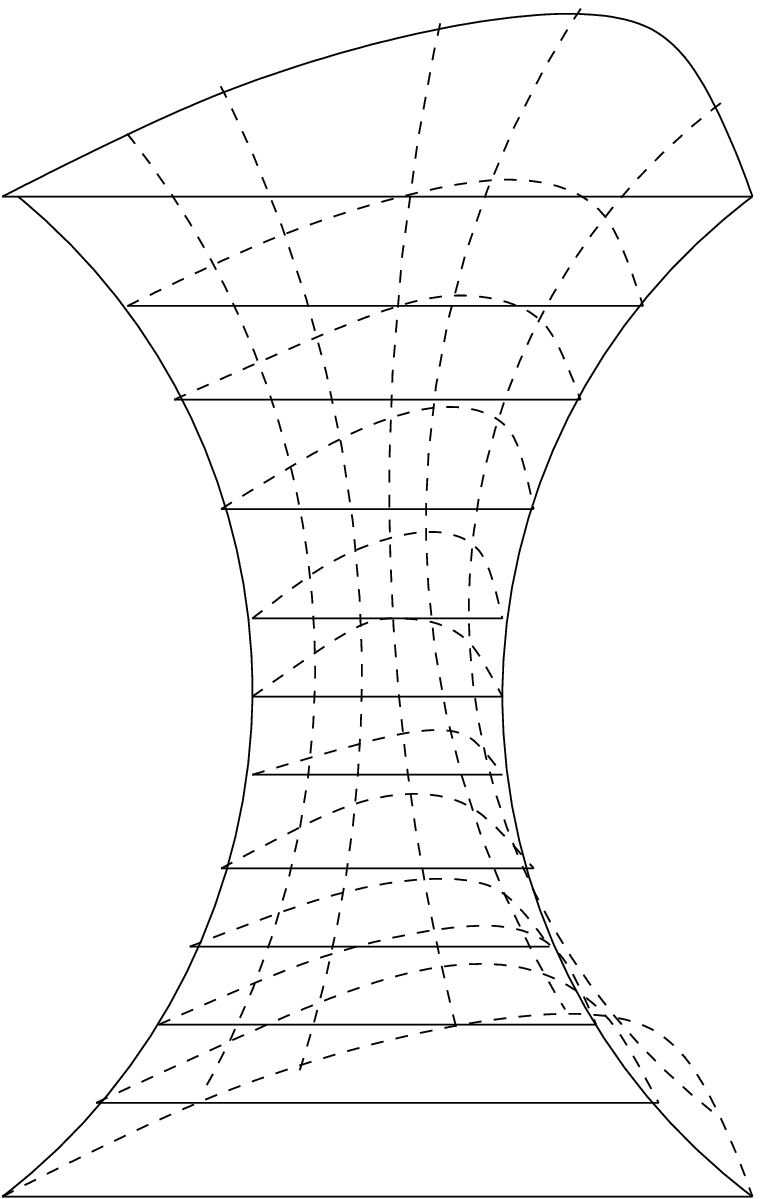} \label{bounceEmbedding}}
\caption{Both the Euclidean and Lorentzian geometries can be embedded
  in higher dimensional flat space. On the left, the Euclidean version
  is the surface of a sphere intersected by a plane. On the right, the
  Lorentzian version is a hyperboloid intersected by a plane. In both
  cases, the domain wall is at the boundary between the curved part
  and the flat part. The cosmological constant jumps across the brane.}
\end{figure}

The Euclidean version of the thin wall geometry can be visualized
by starting with a D-sphere embedded in $D+1$ dimensions. The
sphere is the Euclidean version of de Sitter space. The flat
portion of the space is a D-plane, also embedded in $D+1$
dimensions. Let the plane intersect the sphere as in figure
\ref{euclidean}. The result is the thin wall geometry.

Similarly the Minkowski version begins with a hyperboloid embedded
in $D+1$ dimensional Minkowski space. Intersecting the hyperboloid
with a plane as in figure \ref {bounceEmbedding} yields the thin
wall bounce geometry.

In the above discussion, we have assumed that both the flat region and
the de Sitter region are 4 dimensional. The same framework applies if
the tunneling to flat space entails the decompactification of some
dimensions, so the flat region could be 10 or 11 dimensional.

 \setcounter{equation}{0}
\section{The Causal Patch}

We begin with the concept of a causal patch of pure classical de
Sitter space. Pick a point  $\bf F $ on the future boundary and a
point $\bf P$ on the past boundary. Such a pair of points defines
a causal patch. Now construct the  future light cone of $\bf P$
and the past light cone of $\bf F$. The interiors of these light
cones define the causal future and causal past of  $\bf P$ and
$\bf F$. The intersection of the causal future of  $\bf P$ and the
causal past of $\bf F$ define the causal patch of $\bf{F,P}$. The
intersection of the two light cones is the bifurcate horizon of
the causal patch and the light cones themselves are the future and
past event horizons.

Classical de Sitter space has the symmetry $O(D,1)$, part of which
acts to move the points $\bf P$ and $\bf F$ to new points. In fact
any causal patch of \des \ can be transformed to any other causal
patch by means of this symmetry. Evidently the choice of a
particular causal patch is a {\it gauge symmetry} \cite{fischler}. 
We will return
to this point.

Now consider the quantum version of de Sitter space in
which the boundaries are replaced by quantum fractals of crunches
and hats in both the past and future. Let us pick a point on the
past fractal but not entirely arbitrarily. We choose the point
$\bf P$ to be the tip of a past hat and the point $\bf F$ to be
the tip of a future hat. Otherwise the points are arbitrary. We
believe that even in this quantum case, the choice of points
$\bf{P,F}$ should be viewed as a gauge choice.

A special case of this construction is the causal patch of the
bounce geometry. The regions II and III each have a point at
time-like infinity, as well as light-like infinities. The points
at time-like infinity are the tips of the future and past hats. We
choose the tips of these hats to be the points $\bf P$  and $\bf
F$. The causal patch is the diamond-shaped region shown in figure
\ref{bubbleConfHor}. Note that the causal patch is partly in the
flat domain and partly in the de Sitter domain.

\begin{figure}[!htb]
\center
\includegraphics [scale=.6]
{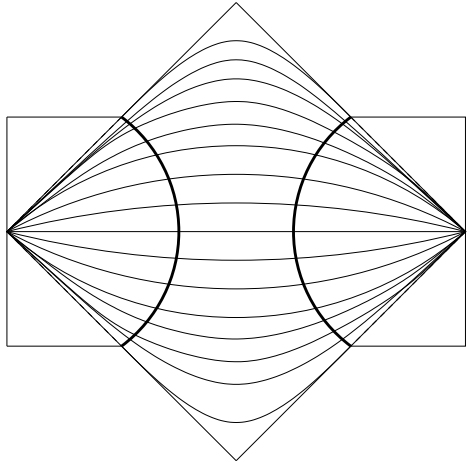} \caption{The causal patch of our observer can be foliated
  by time
 slices
  as shown. The true Penrose diagram is half of the
  figure.}
\label{timeslices1}
\end{figure}

The bifurcate horizon in this case lies at $y=0$ where the local
sphere has vanishing area. The implication of this fact is that
the quantum description of the patch should be in terms
of pure states rather than the entangled states that characterize
de Sitter space or the Schwarzschild black hole. The causal patch
can be foliated with space-like surfaces that cover the whole
spatial geometry, from $y=0$ to $y= \pi$, as in figure
\ref{timeslices1}.

It may seem surprising that an initial contracting FRW solution
does not lead to a singularity. The singularity, if it existed,
would be right at the center of the diagram in figure
\ref{bubbleConfHor}. The ``corners" where the branes end could
easily be interpreted as naked singularities which could create
infinitely energetic light-like shock waves that collide at the
center. But this is not the case. This is made clear by the
Euclidean version of the theory. The Minkowski and Euclidean
geometries agree along the space-like static surface that divides
the diagram in two. Indeed it is possible to define a Hartle
Hawking state on this surface. The origin, where the shock waves
would collide in the Minkowski case is a completely ordinary
nonsingular point in the Euclidean continuation. It is the center
of the flat region of figure \ref{euclidean}. It follows that
there is no singularity in the Minkowski case.

In ordinary de Sitter space the causal patch has a static
geometry. In fact a causal patch is often referred to as a static
patch. But in the bounce geometry, the causal patch is by no means
static. An observer in this region sees a spherical brane that
contracts, bounces and expands. In general the time dependence
will lead to particle creation. In particular as the field rolls
from its initial value at $\phi_b$ to the minimum at zero energy,
particles will be produced.

\bf Particle Distribution \rm

The distribution of particles in the FRW regions (regions II and III of figure
\ref{eternal-all-regions}) is determined by
the symmetry of the solution, namely $O(D-1,1)$ which acts in the
FRW regions on space-like hyperboloids. In the FRW coordinates, the implication
 is that the particle distribution is uniform on the spatial slices. In the embedding
 coordinates, the same symmetry is described as boost invariance, so the
 implication is that
the particle distribution is boost invariant. Such a distribution
will have an infinite number of particles concentrated along the
light-like directions or, equivalently, far out on the hyperboloid.
Since the hyperboloid expands with time, the particle density
tends to zero.

The asymptotic states on the hats have $O(D-1,1)$ symmetry and are
uniquely specified by continuation from the Euclidean theory. One
way to think of it is to use the Euclidean theory to define a
Hartle-Hawking state on the symmetry plane of the solution and
then evolve that state forward and backward to the hats.

Mathematically, a state of this kind should satisfy cluster
decomposition. The way to guarantee this is to begin with the flat
space Fock vacua on the hats. Call these states $|0\rangle_{p}$
and $|0\rangle_{f}$. The correct asymptotic states are obtained by
exponentiating an $O(D-1,1)$-symmetric ``cluster" operator and
applying it to the Fock vacuum. Such cluster operators can be
built by starting with an arbitrary rotationally symmetric
connected operator on a space-like  hyperboloid. Integrating
the position of the operator over the hyperboloid will project out the
$O(D-1,1)$-invariant part of the operator. Note that the
invariance under $O(D-1,1)$ does not determine a unique cluster
operator.  For the case of a non-interacting field theory the
cluster operator is quadratic and the result is a squeezed state.
More generally the clusters are superpositions of any number of
field operators.

The states defined in this way are guaranteed to be nonsingular,
particularly at the origin. But if cluster operators are tampered
with the result will generally lead to a FRW singularity at the
origin. The condition of no singularity is enough to uniquely
determine the $O(D-1,1)$-invariant to be the state that evolves
from the Hartle-Hawking state. We call these states
$|V\rangle_{in,out}$.

The full space of states includes many non-invariant states. In
general perturbing the asymptotic states can lead to
singularities. The potential singularities are due to the infinity
of particles moving out along the light cone. If we trace them
back they all appear to intersect at the origin, i.e. the center
point of figure \ref{bubbleConfHor}. This is obviously a potential
source of trouble. But as long as the boundary conditions far out
on the hyperboloid are unmodified, the particles will be absorbed
by the time dependent field before focusing at the origin. Any
localized perturbation which is made by operators in a bounded
region of the asymptotic hyperboloids should be non-singular. Thus
there is an incoming free particle Hilbert space of states and a
similar outgoing space.

Later we will see that in the quantum version of the bounce there
are nonperturbative processes in which metastable de Sitter space
appears as an intermediate state. In fact it is likely that every
de Sitter vacuum appears as an intermediate state.

\setcounter{equation}{0}
\section{Horizon Complementarity}

The classical geometry of the bounce solution will be modified
by nonperturbative quantum corrections. Among those corrections
are the bubble nucleation events that turn the de Sitter
boundaries into fractal populations of bubble universes. These
bubble universes are on the far side of the event horizon.
According to classical general relativity, events behind the
horizon are completely decoupled from the causal patch and cannot
influence, in any way, observations in the patch.  It has been
argued that this  decoupling of the bubbles from our universe
makes these other universes more metaphysical than physical.

We believe that the complete decoupling is a feature of classical
physics, that does not survive in a complete quantum theory of
gravity. The basis for this belief is the last decade of
experience in understanding black hole horizons, particularly in
the context of string theory. That experience can be summarized by
two principles; The Principle of Black Hole Complementarity \cite{comp}
\cite{fischler}, or
more generally Horizon Complementarity, and the Holographic
Principle (see \cite{holography} for reviews and references). 
Let us review the Horizon Complementarity Principle.

\begin{figure}[!!htb]
\centering \subfigure[]{\includegraphics
[scale=.6, clip]{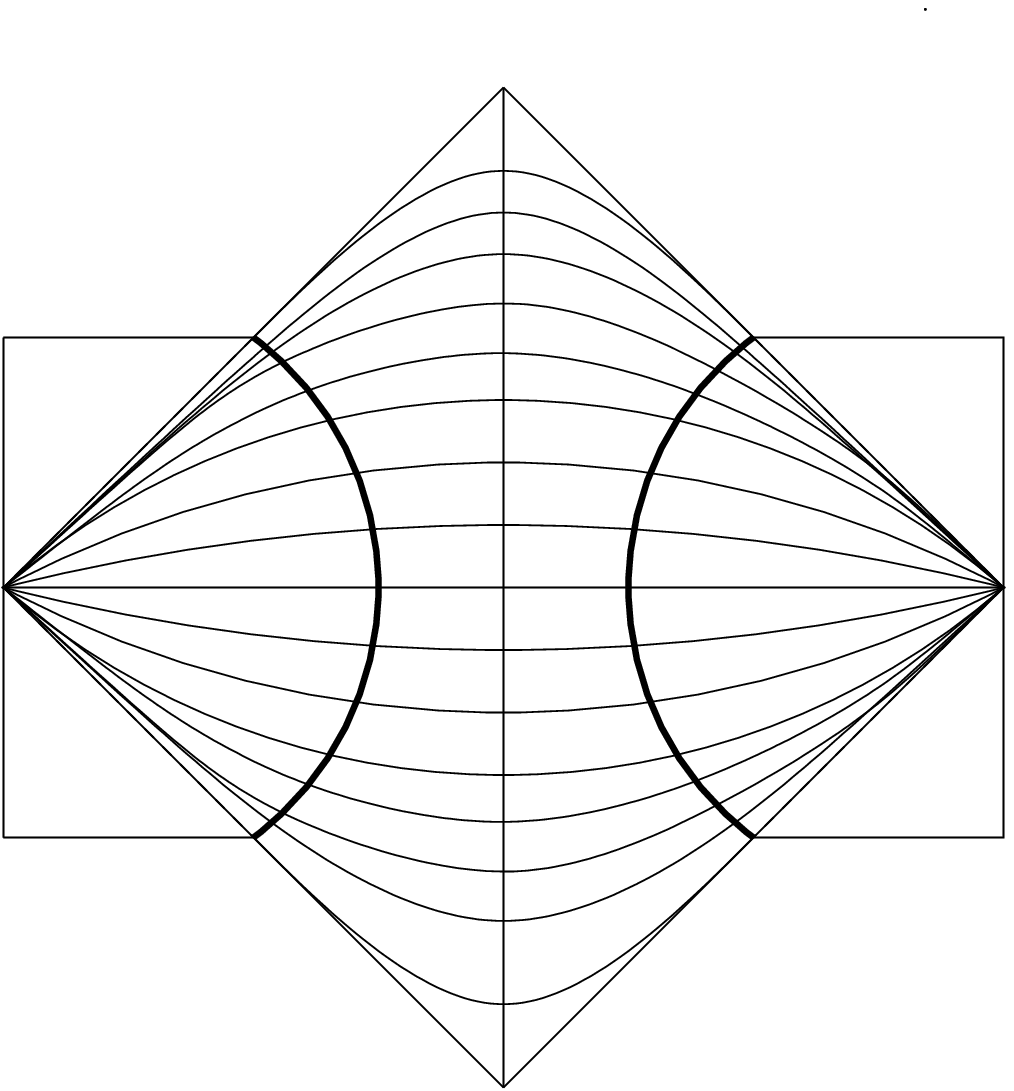} \label{timeslicespen}} \hfill
\subfigure[]
{\includegraphics
[scale=.8]{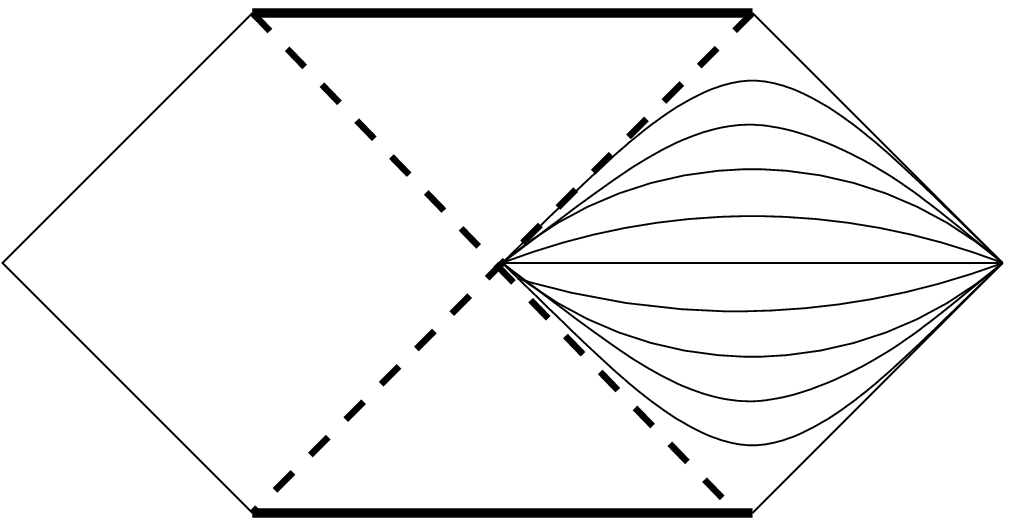} \label{schwarzTimeslices}}
\caption{There are similarities between the Penrose diagram for the
  bounce background (left) and the Schwarzschild black hole
  (right). Constant time surfaces are shown. In both cases, we believe
  that unitary evolution exists between {\it part} of the $t =
  -\infty$ 
surface and {\it part} of the $t = + \infty$ surface.}
\label{timeslices}
\end{figure}

The causal patch can be foliated with a set of space-like surfaces
which all pass through the bifurcate horizon as in figure
\ref{timeslicespen}. Any such set of surfaces allow us to define a
time variable $t$ in the causal patch which runs from $-\infty $
to $+\infty$. Note that the time variable we defined in section
\ref{bounceback} is not suitable because it
 naturally foliates only region I. It is
 generally believed that a self-contained  Hamiltonian description
of physics in a causal patch is possible. Things coming in from
the past horizon or going out through the future horizon are part
of the initial or final conditions at infinite time.

The surface $t=\infty$  is comprised  of two parts. One part is
just the future hat itself which consists of time-like and
light-like infinity. The other portion is not part of the boundary
of the Penrose diagram but defines the future event horizon. The
two regions can be distinguished as follows: Every point in the
Penrose diagram represents a $D-2$ sphere. The area of the local
$D-2$ sphere is finite everywhere in the horizon, but on the hat
it is infinite. Similar things are true for the $t=-\infty$
surface.

Essentially identical things can be said about black hole
geometries. In figure \ref{schwarzTimeslices}  a causal patch of
the Schwarzschild geometry is shown foliated by Schwarzschild time
slices. Again the asymptotic time slice $t=\infty$ consists of a
horizon with finite area and the portion with infinite area, i.e.
light-like and time-like infinity.

The claim that there is a self-contained physics in the causal
patch is not particularly controversial. However there is a much
stronger form of the claim, which for a black hole is called black
hole complementarity. What it basically says is that from the
vantage point of an observer at infinity, \it no information is
stored on the portion of the $t=\infty $ surface with finite area.
\rm In the case of the formation and evaporation of a black hole,
it says that there is an S-matrix connecting the states on the
past asymptotic surface with states on the future surface. The
evidence for this conjecture, in the case of a black hole  is
overwhelming. The bounce geometry is less familiar but we can see
no reason why the same thing should not be true. Thus we postulate
that there is an S-matrix connecting the past hat to the future
hat.

There is an even stronger form of black hole complementarity that
is the real reason for calling it complementarity. It is the
assumption that the degrees of freedom in the Hawking radiation
are redundant descriptions of the degrees of freedom behind the
horizon. According to this formulation, not only do the degrees of
freedom behind the horizon fail to commute with those in the
Hawking radiation, but they can be expressed as functions of the
Hawking radiation variables. The connection is of course extremely
scrambled.

In many cases this strong form of complementarity can be proved
from the weaker form. In these cases the existence of an S-matrix
implies that the degrees of freedom behind the horizon are
redundantly described in terms of the Hawking evaporation
products. The geometry describing the formation and evaporation of
a black hole is an example

\begin{figure}[!htb]
\center
\includegraphics [scale=.8]
{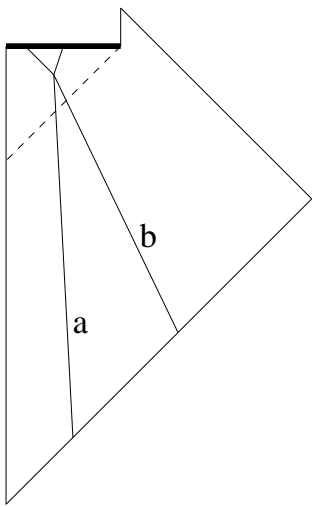}
\caption{Ordinary scattering behind the black hole horizon can be
  translated into a very complicated process occurring outside the horizon.}
\label{bhscatt}
\end{figure}

Figure \ref{bhscatt}  is the standard Penrose diagram for the
formation and evaporation of a black hole. Superimposed on the
diagram is a scattering process in which particles \bf a \rm and
\bf b \rm collide behind the horizon far from the singularity. We
assume that the scattering process is a completely conventional
low energy process such as low energy photon photon scattering.

Suppose we are interested in the outcome of the collision. For
example we might want to know the angle of scattering. Since, in
the in-falling frame, the process is at low energy, ordinary
quantum electrodynamics can be used to describe it. Any observable
can be described by a hermitian operator  including the angle of
the outgoing particles.

Now using the conventional QED Hamiltonian, the observable in
question can be run backwards and re-expressed as an operator in
the Hilbert space of incoming asymptotic scattering states. In
other words by solving the Heisenberg equations of motion, the
observable can be written in terms of operators on past light-like
infinity ${\cal I}^-$. Calling the observable $Q$ \be Q_-=UQU^{\dag} \ee
Where $Q_-$ is an operator on 
${\cal I}^-$ and $U$ is the unitary
operator that connects the final state of the collision to the
incoming states on ${\cal I}^-$. So far all of this can be done using
only conventional low energy physics.

Next we assume the existence of an S-matrix $S$, that connects
states on ${\cal I}^-$ to states on ${\cal I}^+$.
This S matrix is of course not something that we can compute  by
ordinary methods. Among other processes it describes the formation
and evaporation of a black hole.  It scrambles information and
produces the Hawking radiation.

The operator $Q_-$ can now be moved forward in time by conjugating
it with the S-matrix \be Q_+ = S^{\dag}UQU^{\dag}S. \ee The
operator $Q_+$ has exactly the same information as $Q$ but it is
 an operator defined on ${\cal I}^+$--the future null infinity--in the
 products of black hole evaporation.

 This argument is very formal but it does show that the existence of an S-matrix implies that
 the degrees of freedom of the Hawking radiation are a
complementary way of keeping track of events behind the horizon.

In the case of the bounce geometry, the Hawking radiation is
replaced by the particle production in the FRW region due to the
time dependence of the metric and the field $\phi$. Assuming that
the degrees of freedom beyond the horizon are redundantly
described by the infinite sea of particles on the future hat
brings us to a remarkable conclusion: The infinity of bubble
universes are not at all out of contact with our universe. Their
degrees of freedom are all around us in the very subtle
many-particle correlations in the cosmic microwave background. 
Strictly speaking, this conclusion only applies to the final exit
to zero cosmological constant.

\begin{figure}[!htb]
\center
\includegraphics [scale=.8]
{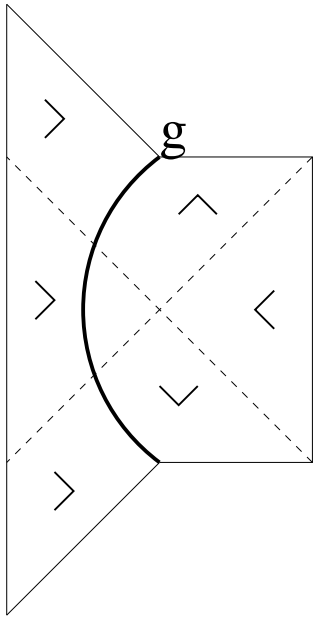}
\caption{There is no bound on the entropy contained in the upper
  ``hat,'' as this Bousso-Penrose diagram shows.}
\label{boussofig}
\end{figure}

The total number of bubbles in the multiverse is expected to be
infinite. This raises the question of whether there are enough
degrees of freedom in the causal patch to describe an infinite
number of bubbles. This question can be answered by calculating
the entropy bound on very late time slices. In figure \ref{boussofig}
the Bousso-Penrose diagram for the bounce geometry is shown. The
maximum entropy on the future hat is equal to the area at point
\bf g \rm and that is infinite. Thus there is no bound on the
amount of information that can be stored on the upper (or lower)
hat \cite{bousso}.

\bf The Conjecture \rm

The conjecture that we would like to put forward is this:

1. Asymptotic in and out ``vacua", invariant under $SO(D-1,1)$
exist and lead to nonsingular behavior in the bulk. The vacua
have an infinite number of particles when expanded in a flat space
basis of states.

2. A Hilbert Space of asymptotic in and out states consisting of
localized perturbations on the respective vacua exist.

3. The asymptotic in and out states are connected by an S-matrix.

The particle density in the asymptotic vacua vanishes because the
FRW expansion dilutes the particles. This means that the
asymptotic vacua are locally identical to some conventional flat
space vacuum with zero cosmological constant. The only such vacua
that we know of are on the moduli space of supersymmetric vacua of
string theory. Thus we assume that the vacua on the hats
are always supersymmetric, at least locally.
Different metastable vacua may lead to different points on the
supersymmetric moduli space.

\setcounter{equation}{0}
\section{Perturbations around  the Bounce Background}

\subsection{The equations of motion \label{eqmot}}

Ultimately we would like to show that string theory in the bounce
geometry exists and can be used to calculate the S-matrix. We are
far from that goal but there are a number of interesting issues
which are likely to come up. Let's for the moment forget string
theory and think about quantum field theory in the bounce
background. Presumably Feynman rules can be constructed for the
perturbative processes. The obvious way to carry out a
perturbation theory would be to start with the Green functions in the
Euclidean version of the theory and continue them to the Minkowski
signature. Then by means of the LSZ technique, an S-matrix can be
constructed.

In Appendix A, we argue that quantum field theory is well-defined
in the bounce background, explain some subtleties due to working
on a compact space, and compute some correlation functions. In the
main text, we focus on the nonperturbative processes because they
are the most interesting for our purposes. We certainly don't have
a systematic formulation but some idea of the nature of these
processes can be gained by a kind of mini-superspace reduction of
the theory. For simplicity we will work in the thin wall
approximation, but the general case is a straightforward
generalization.

Although the entire causal patch cannot be described in static
coordinates with a time-like Killing vector, region I of the
Penrose diagram does have a static description. For our present
purpose this is sufficient. First consider the static metric for
the pure bounce solution. The radial coordinate $r$ is defined to
be zero at the origin of the flat space region.

We want to describe a spherically symmetric bubble of $\Lambda =
0$ inside a region with positive $\Lambda$.  We will closely follow \cite{guth}.
In the simplest
situation, the \0cc\ region is pure Minkowski space and the \pcc\
region is pure de Sitter space. If we use static coordinates on
both sides, then the metric on each side has the form \be ds^2 =
-f(r) dt^2 + {1 \over f(r)} dr^2 + r^2 d\Omega^2,
 \label{metric}
\ee with $f(r) = 1$ in the flat part and $f(r) = 1- r^2/ R^2$ in
the de Sitter region, where $R$ is the de Sitter radius.

We will study two spherically symmetric perturbations  of this
geometry. One possibility is to add a mass $M$ at the
center of the flat region, $r=0$, so that the metric inside the brane
is Schwarzschild and the metric outside remains pure de Sitter
space. Another possibility
is to add a mass $M$ at the center of the de Sitter region, 
so that the metric is Schwarzschild-de Sitter 
on the \pcc\ side and Minkowski space on the
\0cc\ side. To be clear, the time slices are spheres cut by planes 
just like figure \ref{euclidean} and
we add a mass at the point on the sphere farthest from the plane.
In both cases, the gravitational backreaction changes the
motion of the brane, so we get a one-parameter family of brane motions.
Let us
describe the brane motion by $r(\tau)$ where $\tau$ denotes proper time
along the domain wall trajectory and $r$ is the Schwarzschild radial 
coordinate.

First we discuss adding a mass in the de Sitter region.
In equations, the effect of adding a mass in the \pcc\ region is that
for $r< r(\tau)$ the metric is \be ds^2= - f_{in} dt^2 +
{1\over f_{in}}dr^2 + r^2 d\Omega^2 \label{schw} \ee with \be
f_{in}= 1. \label{fin} \ee
The metric for $r> r(\tau)$ is \be ds^2= -
f_{out} dt^2 + {1\over f_{out}}dr^2 + r^2 d\Omega^2 \label{desit}
\ee with \be f_{out}=(1-{r^2 \over R^2} - {2 G M \over r}). \label{fout} \ee

To determine the domain wall position $r(\tau),$ we need to know
what equation of motion it satisfies. The Israel junction
condition gives the equation of motion for the brane. Given the
Schwarzschild-like form of the metric on both sides, the condition is
\be \sqrt{f_{in}(r) + \dot r^2} -
\sqrt{f_{out}(r) + \dot r^2} = 4 \pi G \sigma r,
\label{junctioncondition} \ee where $f_{in}(r)$ is the metric
function inside the bubble, $\dot r$ is the derivative of the
Schwarzschild radial coordinate with respect to proper time, and
$\sigma$ is the tension of the domain wall.

The junction condition can be rearranged to look like an energy
conservation equation for the brane motion,
\be 4\pi \sigma r^2 \sqrt{1 + \dot r^2}-({1 \over 2 G R^2} + 8\pi^2 G
\sigma^2)r^3 = M. \label{meq}\ee
Each term has a physical interpretation. The square root term is the
usual kinetic term for a membrane, and $-8 \pi^2 G \sigma^2 r^3$ is the
gravitational self-energy of a spherical membrane.  The bubble of flat
space replaces a region of positive vacuum energy with a region of
zero vacuum energy, resulting in a change
$-{1 \over 2 G R^2}r^3$ in the energy.

Why should the parameter $M$, which we thought of as a mass in the de
Sitter region, have a nice interpretation as the energy of the bubble
of flat space? Roughly, it is because nonsingular perturbations of de
Sitter space involve adding equal masses on opposite sides of the
spatial sphere. When we add a mass at the north pole of the spatial
sphere, the gravitational backreaction adjusts the position of the
brane so that the bubble of flat space effectively has positive energy.

\begin{figure}[!htb]
\center
\includegraphics [scale=.8]
{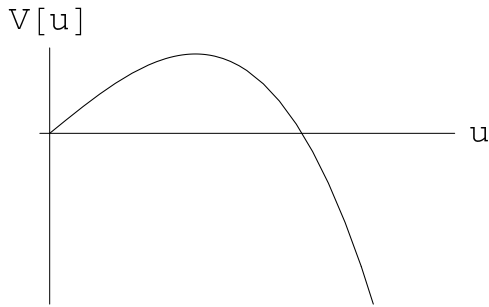}
\caption{The effective potential for the motion of the domain wall has
a barrier. The classical evolution corresponds to coming in from
infinity, bouncing off the barrier, and going back out. Tunneling in
to the center means that the brane reaches 0 radius: it
disappears, leaving pure de Sitter space.}
\label{dsbubblePotential}
\end{figure}

Equation \ref{meq} is intuitive, but since we want to analyze the
one-dimensional problem we will make it look like the energy
conservation
equation for a
particle rather than a brane.
Take the non-relativistic limit ${\dot r} <<1$ and change variables to
$u = r^2$. The equation becomes
\be
\12 \pi \sigma \dot{u}^2 + 4 \pi \sigma u - ({1 \over 2 G R^2} +8 \pi^2 G)
u^{3/2} 
= M,
\ee
which has the form an energy conservation equation
for a non-relativistic particle in a potential. The energy is
given by $M$ and the potential is 
\be
 V(u) = 4 \pi \sigma u - ({1 \over 2 G R^2} +8 \pi^2 G)u^{3/2}
\ee
The potential is shown in figure \ref{dsbubblePotential}. We will
discuss the interpretation further in section \ref{reson}.

A similar one-dimensional problem may be constructed for the family of
solutions obtained by adding a mass at the center of the flat
region.
The boundary conditions are that outside the bubble, $r>
r(\tau)$, the geometry is pure de Sitter while inside it is given
by the metric of the massive particle in flat space. For $r<
r(\tau)$ we take the metric to be Schwarzschild. We assume that
the Schwarzschild radius of the mass $M$ is much smaller than the
other length scales in the problem. Now
\be
f_{in}=(1-{2MG \over r}) 
\ee
and
\be f_{out}=(1-{r^2 \over R^2}). \ee
The junction
condition  can be rearranged to the form
\be {4\pi \sigma r^2 \sqrt{1 - r^2/R^2 + \dot r^2} \over
  \sqrt{1-r^2/R^2} } -({1 \over 2 G R^2} - 8 \pi^2G
\sigma^2)r^3 = -M. \label{nonrel} \ee

To make this look like a conventional energy conservation equation for a
particle, again let $u = r^2$ and take the limit where the
minimum brane size is much smaller than the de Sitter radius, $G
\sigma << 1/R$, so that the nonrelativistic motion occurs in the
region $r << R$. The equation of motion becomes
\be
\12 \pi \sigma \dot u^2 + 4 \pi \sigma u - {1 \over 2 G R^2} u^{3/2} =
-M.
\ee
The potential 
looks just like figure \ref{dsbubblePotential}, but now positive mass
corresponds to {\it negative} energy.

\begin{figure}[!!!!htb]
\center
\includegraphics [scale=.8]
{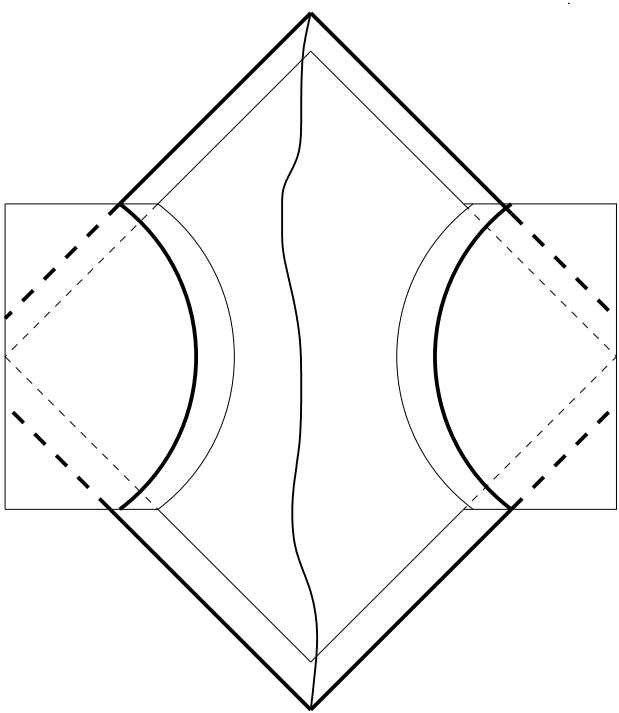}
\caption{Adding a mass inside the bubble causes the causal patch of
  the observer at the center to ``expand'' as shown in the figure. The
  new causal patch is shown in bold.}
\label{massInside}
\end{figure}

It is not obvious what happens to the causal structure for either of
these perturbations. The question is important because the
bounce solution is on the verge of having a horizon. In Appendix B, we
show that adding a mass at the center of the flat region causes the
brane to move out from the origin, as
shown in figure \ref{massInside}, so no horizon forms. We believe that
adding a mass at the center of the de Sitter region has exactly the
opposite effect.

\subsection{de Sitter as a Resonance \label{reson}}

For $M=0$ there are two classical solutions. One of them is just
the original bounce solution. The other is the static solution
located at the minimum of the potential at $r=0$. The
interpretation of that solution is especially interesting. It
represents the spherical brane of zero radius-- in other words, no
brane at all. This solution is just pure unmodified de Sitter
space!

In the quantum theory the  static solution with the degenerate
vanishing domain wall is unstable; it can tunnel out through the
barrier. This is just the instability of de Sitter space to bubble
nucleation.

For the program of defining and studying an S-matrix, the state at
$r=0$ -- the pure de Sitter space-- is an intermediate resonance, a
singularity as a function of energy. The singularity is not
exactly at zero energy because the finite lifetime of the state
shifts the singularity into the complex plane.

In a semiclassical analysis of the S-matrix we can distinguish two
types of histories. The first consist of small fluctuations around
the bounce solution. These can be studied using conventional
perturbation theory in the bounce background. The second type of
history involves the formation and decay of the resonant pure de
Sitter intermediate states. These histories begin with an incoming
solution identical to the bounce solution. The domain wall moves
inward until the point where it comes to rest. This occurs at the
point \be r={8\pi G \sigma \over 16\pi^2 G^2 \sigma^2 + R^{-2}}.
\label{minr} \ee

\begin{figure}[!htb]
\center
\includegraphics [scale=.8]
{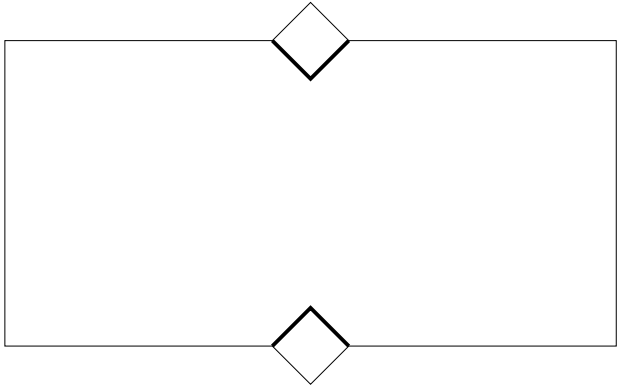}
\caption{A ``solution'' in which the brane tunnels into the minimum at
zero size for a time can be illustrated by this conformal diagram. The
evolution is classical except at the points where the brane disappears
and is nucleated. It is not really nucleated at zero size unless the
tension is zero.}
\label{zerotensiontunn1}
\end{figure}

At this point the fictitious particle tunnels to the origin where
it remains for a time $t$. It then tunnels back to the point
(\ref{minr}) and continues on its original course with a time delay
$t$. The process can be thought of as an interrupted bounce. It
can be illustrated by the conformal diagram shown in figure
\ref{zerotensiontunn1}.

The amplitude for this process has the form \be A= \gamma
\int_0^{\infty} dt \label{amp} \ee where $\gamma$ is the
  Coleman de Luccia tunneling rate given by
\be \gamma \sim e^{-S }. \ee Here $S$ is the action for the
Euclidean bounce solution. From (\ref{amp}) the amplitude appears to diverge. To see the
meaning of this, let us shift $M $ away from zero. In that
case the integrand picks up an additional factor $e^{iMt}$ and the
result of the integration is \be A=\gamma {1\over M}. \ee

The pole at $M=0$ is the standard pole in the energy indicative of
a sharp intermediate state. However in a more precise calculation
including re-scattering corrections the pole associated with an
unstable state gets shifted into the complex plane. Thus the
amplitude will have the approximate form \be A=\gamma {1\over M
+i\gamma}. \label{respol} \ee
In Appendix C, we give a more careful argument for the same result
which does not rely on using the non-relativistic approximation or
defining the energy via the mass of the added particle, but the result is
exactly the same.

In fact the pure de Sitter space is not a single resonant state.
It consists of a large number of nearby states implied by the
thermal density matrix describing the quantum mechanics of de
Sitter space. Although, like any unstable state, the de Sitter
states are not well-defined quantum states, they should have
precise meaning as resonances in the complex plane. In this
respect they are like black holes.

Although our goal is the description of non-supersymmetric
metastable de Sitter vacua, supersymmetry plays a central role.
The entire possibility of an S-matrix description depends on the
existence of vacua with exactly zero cosmological constant. There
is no reason to expect such vacua in the absence of supersymmetry.
Indeed the S-matrix elements described in this paper are a sector
of the S-matrix of a theory whose asymptotic states are classified
by supersymmetry.

\subsection{Bubble Collisions \label{bubcol}}
Up till now we have ignored the possibility of collisions between
bubbles. These are not only possible but are inevitable. To see
this recall that along any time-like trajectory, an observer
eventually is swallowed by a bubble nucleation event. Consider a
trajectory that eventually ends at a point where the domain wall
meets the future hat. Classically an observer following such a
trajectory sees herself in de Sitter space. Eventually she will
encounter a bubble nucleation. Obviously the expanding bubble will
collide with the original bubble. This process will occur an
infinite number of times.

One might think that the bubble collisions would have the effect
of connecting all the bubbles together into one big bubble that
covers the whole space. But this is not correct. Guth and Weinberg
analyzed this situation in \cite{gw}. They find
that the bubbles cluster into disconnected structures that
maintain their island-like character.

\begin{figure}
\center
\includegraphics [scale=.5]
{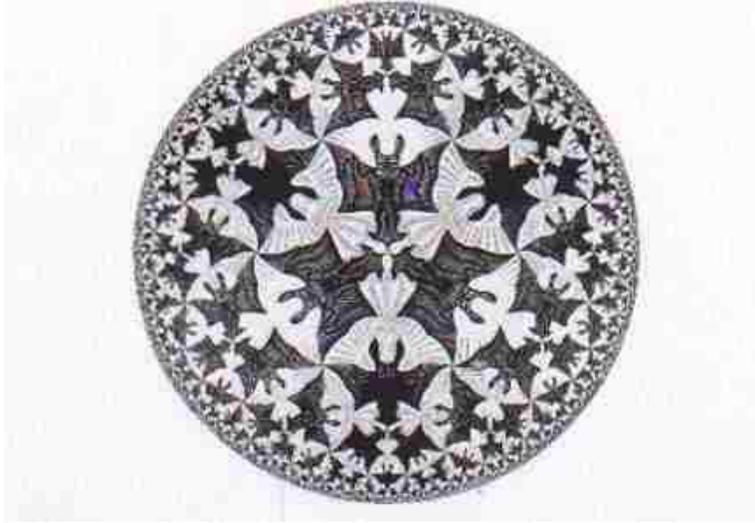}
\caption{Escher's famous figure illustrates properties of the
  hyperbolic plane, which is the spatial geometry seen by the FRW observer.}
\label{escher}
\end{figure}

Bubble collisions certainly make nonperturbative corrections to
the final state of the hat. But because the entire cluster forms
in a background with $O(D-1,1)$ symmetry, the final state must
have this invariance. This means that the perturbations caused by
such collisions are uniformly distributed over the  hyperbolic
plane representing a spatial slice through the FRW region. In
figure \ref{escher} we show one of Escher's drawings
of the two dimensional hyperbolic plane tessellated by ``Angels and
Devils". The different bubbles that coalesce to form the cluster
should be distributed like the devils in the figure,
although much more sparsely.

Since the final states are expected to lie on the supersymmetric
moduli space, the different patches will eventually settle down to
give an inhomogeneous state with varying massless scalar fields
distributed symmetrically over the negatively curved geometry. The
background field is given by the average field and the variations
are part of the distribution of massless particles in the final
state. If this picture is correct then bubble collisions do not
destroy the overall picture but they do give nonperturbative
corrections to the particle content on the hat. We hope to return
to this point in the near future.

\setcounter{equation}{0}
\section{Conclusion}

In this paper we have suggested a framework for studying eternal
inflation within the context of an S-matrix description that may
be adaptable to string theory. The metastable de Sitter vacua
appear as resonant intermediate states in the scattering matrix
elements.

This framework, when combined with horizon complementarity, leads to
a conceptually far-reaching conclusion: The infinity of bubble
universes out beyond an observer's horizon is not truly
decoupled. Radiation analogous to Hawking radiation exists and
encodes the degrees of freedom of these other bubbles in a
scrambled way. This radiation is essentially the cosmic microwave
background, but because our universe still has a non-vanishing
cosmological constant there cannot presently be enough
information to encode all the other bubbles. It's interesting to
quantify this a little more. Given the current value of the
cosmological constant, the entropy of our horizon is about
$10^{120}$. Most of this is in the horizon degrees of freedom. The
amount of non-horizon entropy in ordinary stuff is about
$10^{100}$. This means that the horizon degrees of freedom contain
enough information to describe the features of $10^{20}$ universes
like our own.\footnote{W. Fischler has pointed out to us that it is misleading to say this information is stored in the CMB. The characteristic wavelength of the Hawking radiation is horizon size, so it is the far infrared of the CMB. Additionally, the time scale to extract information from the Hawking radiation is extremely long, just as it is for black holes. For a black hole, the time to extract one bit of information is $r^3/l_p^2$, where $r$ is the Schwarzshild radius. Assuming that for de Sitter space we should replace the Schwarzshild radius by the de Sitter radius, the time is $10^{120}$ times the age of the universe.}

But if we wait long enough our universe will tunnel to an open FRW
bubble with an infinite number of particles. Those observers who
survive the transition will have enough available information to
reconstruct the rest of the universe.

The reason we emphasize this point is not to suggest that there is
a practical way of testing the hypothesis of eternal inflation.
Even if we could wait long enough to enter the FRW era and collect
enough quanta, the information would be in a hopelessly scrambled
form. Our  motivation is to dispel the idea that discussing the
portion of the universe beyond our classical horizon is pure
metaphysics. We would argue that the rest of the universe will
become imprinted in the causal patch and is physically meaningful.

Another important question is the unitarity of the S-matrix. The
initial and final vacuum states are determined by the background
de Sitter vacuum that the bouncing bubble is embedded in. If the
de Sitter vacuum is replaced by another minimum in the Landscape,
the boundary conditions in the FRW spatial infinity will change.
Moreover it seems possible that transitions can occur between
these asymptotic vacua. Consider a history involving a tunneling
to  a particular de Sitter intermediate state. The system can then
tunnel back to the original state or it might tunnel to some other
nearby de Sitter minimum. If it does the latter it can
subsequently tunnel to a different final state. Thus it seems that
transitions between asymptotic vacua are possible. If this is the
right idea then the S-matrix would only be unitary after summing
over the different sectors with different boundary conditions.

We think that this may not be the right idea. To see why, consider
pure de Sitter space. In this case the choice of points $\bf P$
and $\bf F$ is obviously a gauge choice. Now consider the real
situation in which the initial and final boundaries are replaced
by  fractals containing an infinite number of hats. In fact one
can expect an infinite number of hats with every possible boundary
condition. But if we are right about the implications of horizon
complementarity, then each hat contains all the degrees of freedom
of the other hats. That suggests that the choice of hat in the
initial and final state is a \it gauge choice. \rm  One may pick a
gauge in which the boundary conditions are some specific point on
the moduli space or a different gauge with a different asymptotic
behavior. Some choices of gauge may be more convenient than
others, making manifest the physics with the particular asymptotic
behavior. The physics of other gauges will only exist in a
scrambled form.

If this latter view is right then the S-matrix with a given
asymptotic vacuum may be unitary by itself.

Finally let us address the issue of formulating string theory in
bounce backgrounds. If the bounce were truly a classical string
solution, the formulation of string theory would be
straightforward. But it seems certain that the existence of de
Sitter minima is
 a nonperturbative feature of string theory. As an example, the KKLT model 
 \cite{kklt} relies on
 nonperturbative instanton effects to stabilize the Kahler moduli.

 This does not necessarily mean that string theory cannot be formulated in this kind of background,
 but it does mean that some nonperturbative version of the Fischler-Susskind mechanism will be needed
to define the string action.

\section{Acknowledgements}
We would like to thank Tom Banks, Stephen Shenker, Eva Silverstein,
Lukasz Fidkowski, Andrei Frolov, Xiao Liu, Matt Kleban, Simeon
Hellerman, Darius Sadri, Liam McAllister, Boaz Nash, Veronika Hubeny,
Alex Maloney, M.M. Sheikh-Jabbari, and Atsushi Higuchi 
for helpful discussions. We especially want to thank Vincent Moncrief 
for repeatedly explaining linearization instabilities to us. We thank Steve Giddings and Willy Fischler for suggesting revisions to our first version.

\begin{appendix}
\setcounter{equation}{0}
\section{Perturbation Theory in the Bounce Background}
Perturbation theory in the bounce background is well-defined, but
it has several unique features compared to ordinary flat-space
perturbation theory.  One difference is that the spatial slices
are compact, which leads to interesting restrictions on the
allowed perturbations. Another difference is that an infinite
number of particles are produced. These complications make
perturbation theory more confusing, but no less well-defined.
We first compute some simple correlators and discuss the
particle production. Then we deal with subtleties due to the
unusual geometry of the background.
%


\subsection{Correlation Functions in the Bounce Background: Overview and
  computation of particle production}

To begin, we will compute the correlator of a conformally coupled
scalar field in the thin-wall approximation. In spite of the time
dependence of the background, the correlators in the \0cc\ region
are exactly the same as they would be in ordinary flat space!

The reason for this is that we can conformally map the Euclidean version
of the bounce, shown in figure \ref{euclidean}, to the flat plane using
a conformal mapping which is trivial in
the flat region.

It may be surprising that the complicated geometry has no effect on
correlators in the flat region.
A helpful analogy is the
uniformly accelerated mirror in flat space. The correlators of a
massless field in flat space are unaffected by the presence of the
mirror {\it if} the acceleration is uniform \cite{b&d}.

{\it The Conformal Mapping.} The Euclidean space consists of a
sphere sliced by a plane. For simplicity, we consider the special
case where the plane slices the sphere exactly in half. We
coordinatize the flat part in the standard way. The half sphere we
coordinatize by stereographic projection from the opposite pole
onto the plane.  The metric is 
\bea
ds^2 & = & dr^2 + r^2 d\Omega^2, \mbox{ for } r < 1 \nonumber\\
ds^2 & = & \frac{1}{(r^2 + 1)^2}(dr^2 + r^2 d\Omega^2), \mbox{ for } r > 1
\eea 
In these coordinates, it is clear that the space is
conformally flat both inside and outside the domain wall. The
conformal factor is continuous at the domain wall, so the space is
conformally flat everywhere.
A subtlety is that the correlator of a
conformally coupled field is not well-defined. This is easiest to
see in two dimensions, where a massless field is conformally
coupled. The shift symmetry $\phi \rightarrow \phi + c$ ensures
that the correlator is undefined. Correlators of derivatives {\it are}
well-defined.

A hidden assumption in this section is that the whole effect of the
 domain wall
on the field $\phi$ is through the geometry. In other words, we implicitly
imposed continuity of the field and its derivative at the domain
wall rather than a more complicated boundary condition.
 
\subsection{Subtleties due to compactness and symmetry of the
 background \label{compsymm} }
It is well known that the total charge on a compact space must be
zero, roughly because there is no place for electric field lines
to end except on charges. Since the bounce solution has compact
spatial slices, we must restrict to
electrically neutral perturbations. There is a less familiar
gravitational analog of the restriction on total charge which we will describe.

First, consider a familiar example from string theory. When
quantizing the closed bosonic string in lightcone gauge, one
imposes various gauge conditions on the worldsheet metric. These
conditions completely fix the worldsheet reparameterization
invariance except for the transformations \be \sigma \to \sigma +
const \ee where $\sigma$ is the periodic coordinate on the string
worldsheet. The unfixed gauge freedom means we must impose a
condition on the physical states, restricting to states which are
invariant under the residual gauge transformation. Since the gauge
freedom is translations along the string, we restrict to states
with zero worldsheet momentum. This restriction is the
familiar level matching condition.

To state the condition in a form which will generalize, the
residual coordinate transformation is generated by the Killing
vector \be \xi = {\partial \over \partial \sigma } = {\partial
\over \partial z} - {\partial \over \partial \bar{z}} \ee This
leads to the constraint \be 0 = \int d\Sigma \ T_{a b} \ \xi^a
\ \hat{n}^b, \ee where the integral is taken over a spatial slice,
$T_{a b}$ is the worldsheet stress tensor, $\xi^a$ is the Killing vector,
and $\hat{n}^b$ is the
unit normal vector to the spatial slice. In traditional notation,
the constraint is \be 0 = \int d\sigma (T_{zz} - T_{\bar{z}
\bar{z}}), \ee which is the level-matching constraint.

Analogous constraints appear in the bounce solution. Recall that
the gauge invariance for graviton fluctuations is \be h_{a b} \to
h_{ab} + D_a \xi_b + D_b \xi_a. \ee If $\xi$ is a Killing vector
field, then by definition it satisfies \be D_a \xi_b + D_b \xi_a =
0 \ee so it does not generate any gauge transformation. As a
result, the graviton gauge fixing leaves unfixed the coordinate
transformations generated by the Killing vectors of the
background. It is easiest to think about in the Euclidean
geometry, figure \ref{euclidean}. The geometry is invariant under
rotations of the $(D-1)$-sphere. In the Lorentzian version, some
of these rotations become boosts. If we use coordinates so that
the bounce is a hyperboloid cut by the plane $x = const$, then the
Killing vectors are rotations and boosts which leave $x$ fixed.
The corresponding quantities which are set to zero are once again
\be 0 = \int d\Sigma \ T_{a b}\  \xi^a \ \hat{n}^b, \ee where the
integral is again over a spatial slice. The
constraints coming from the rotation generators are, for example,
\be 0 = \int d\Sigma (y T_{0 z} - z T_{0 y}) \ee which sets the
angular momentum in the $yz$ plane equal to zero. (We continue to
use the embedding coordinates because the symmetries are clearest
there.) The constraints coming from boost generators are
 \be 0 = \int d\Sigma \  y \ T_{00} \ee 
if we choose to integrate over the spatial slice $x_0 = 0$.
These constraints are
less familiar, setting the dipole moment of the energy
distribution equal to zero in the symmetry directions. Just as
angular momentum is the charge associated to rotations, the dipole
moment of the mass distribution is the charge associated to boosts
and it should be set to zero.

Why aren't there analogous restrictions in flat space, which after
all has many Killing vectors? The difference is spatial
compactness. In flat space, the modes which are unfixed by the
gauge conditions are not normalizable, so we do not need to impose
their equations of motion as constraints on the physical states.

The analogy with the level-matching condition is incomplete because
4-dimensional gravity has propagating degrees of freedom while 2-dimensional
gravity does not. At higher order in perturbation theory, there cannot
be a restriction on the matter stress-energy tensor, roughly because there is also
stress-energy in gravitational waves. Another way to see the same
thing is to note that the quantity we are setting to zero, 
$\int d\Sigma \ T_{a b}\  \xi^a \ \hat{n}^b$, is not gauge invariant.

Once we allow propagating gravitons rather than just computing the backreaction, we may as well discuss pure gravitational perturbation theory with no matter since no new issues arise when adding matter. In pure gravitational perturbation theory, the combination of compact spatial slices and symmetries leads to a {\it linearization instability}. We thank Vincent Moncrief for explaining the situation to us and refer the reader to his review article \cite{moncrief}, which we follow here, for a more thorough description and references to the literature. The bottom line is that the linearized equations of motion have spurious solutions which cannot be continued to higher order; they are not approximations to any solution of the full equations of motion. To be clear, there is no instability of the background. What is unstable is the linear approximation to the equations of motion for small fluctuations about the background.

 A simple example of a situation where some solutions to the
 linearized equations cannot be continued to higher order is a cone
 defined by $x^2 + y^2 - z^2 = 0$.  Say we perturb around the point
 $(0,0,0)$ and ask which points are on the cone. The point
 $\epsilon(a,b,c)$ satisfies the cone equation to first order in
 $\epsilon$ for any $(a,b,c)$, so it appears that we can move in 3
 directions and stay on the cone. We know this is a wrong result; the
 linearized equations do not give an accurate picture of the space of
 solutions. If we blindly computed in perturbation theory, at second
 order we would find a quadratic constraint on the first order
 fluctuations, namely $a^2 + b^2 - c^2 = 0$. The true linear
 approximation to the theory near such a conical point is the
 linearized equations of motion {\it plus} an extra constraint
 quadratic in the fluctuations.
 
 In the gravity problem, the basic result is that solutions with compact spatial slices and Killing vectors are conical points in the space of solutions. Just as in the cone example, there are solutions of the linearized equations of motion which are not approximations to any full solution. The true linear approximation consists of the linearized equations of motion plus a second-order condition \cite{moncrief}
\be
\int d\Sigma \ G_{a b}\  \xi^a \ \hat{n}^b
\ee
where $G_{a b}$ is the Einstein tensor to second order in the metric fluctuations. There is one such condition for each Killing vector.

To summarize, there are some extra subtleties in doing gravitational
perturbation theory in the bounce background, but these subtleties are
well understood by relativists and constitute an inconvenience rather
than a disease. We have not yet determined the most convenient way to
compute in such a background.\footnote{Pure de Sitter space is the
simplest case of a background with linearization
instabilities. Higuchi and Weeks 
\cite{higuchi} have computed the graviton propagator in de Sitter
space without worrying about the linearization instability. As suggested by Moncrief, one could possibly use their propagator and impose the extra constraints on the states. We have not yet pursued this possibility.} At the moment our main goal is simply to establish that perturbation theory makes sense in the bounce background.

\setcounter{equation}{0}
\section{Motion of Domain Wall with Mass Inside}
Here is the computation to support the claim of section \ref{eqmot}
that adding a mass inside the bubble causes the
domain wall to move out so that there is no horizon at all. Begin
with the junction condition 
\be \sqrt{1 - 2GM/r + \dot r^2} - \sqrt{1 - r^2/R^2 +
\dot r^2} = 4 \pi G \sigma r \ee 
We will deal with small $M$. At
the time when $\rd = 0$, the radius of the bubble is given by
$r_0$, where $r_0$ satisfies \be \sqrt{1 - 2GM/r_0} - \sqrt{1 -
r_0^2/R^2 } = 4 \pi G \sigma r_0 \ee We will expand around a
solution with $M = 0$ which has the same minimum size. We think of
this solution as having a different tension $\sigma_1$: \be 1
- \sqrt{1 - r_0^2/R^2 } = 4 \pi G \sigma_1 r_0 \ee For small
$M$, subtracting the two equations shows that \be \label{sigmaeq}
4 \pi G r_0 (\sigma_1 - \sigma) = G M/r_0 \ee It is more
convenient to think of $\tau$ as a function of $r$ rather than
vice versa. We call the unperturbed solution $\tau(r)$ and the
perturbed solution $\tt(r)$. Derivatives with respect to $r$ are
denoted by primes. Then $\tt(r)$ and $\tau(r)$ solve \bea
\label{taueq}
\sqrt{1 - 2GM/r + 1/\tt^{\prime 2}} - \sqrt{1 - r^2/R^2 + 1/\tt^{\prime 2}} = 4 \pi G \sigma r \\
\sqrt{1 + 1/\tau^{\prime 2}} - \sqrt{1 - r^2/R^2 + 1/\tau^{\prime
2}} = 4 \pi G \sigma_1 r  \label{radeq} \eea Expanding $\tt(r) = \tau(r) +
\epsilon(r)$, these equations become \bea \sqrt{1 - 2GM/r +
1/\tau^{\prime 2} - 2 \epsilon^\prime/\tau^{\prime 3}} - \sqrt{1 -
r^2/R^2 + 1/\tau^{\prime 2}
 - 2 \epsilon^\prime/\tau^{\prime 3}} = 4 \pi G \sigma r \\
\sqrt{1 + 1/\tau^{\prime 2}} - \sqrt{1 - r^2/R^2 + 1/\tau^{\prime
2}} = 4 \pi G \sigma_1 r \eea Subtracting the two equations,
expanding the square roots for small $\epsilon$ and small $M$, and
using the formula (\ref{sigmaeq}) to eliminate the tensions, we
get \be - \frac {GM} { r \sqrt{1 + 1/\tau^{\prime 2}}} - \frac
{\epsilon^\prime} {\tau^{\prime 3} \sqrt{1 + 1/\tau^{\prime 2}}} +
\frac {\epsilon^\prime}{\tau^{\prime 3} \sqrt {1 - r^2/R^2 +
1/\tau^{\prime 2}}} = - {G M r \over r_0^2}  \ee Simplify this by using the
equation satisfied by $\tau(r)$, (\ref{radeq}), which is more
useful in the form \be \label{doteq} 1 + 1/\tau^{\prime 2} = r^2/r_0^2, \ee
to get \be G M r_0/r^2 + \epsilon^\prime (r^2/r_0^2 -
1)^{3/2}(\frac {r_0} r - \frac {r_0} {r \sqrt{1 - r_0^2/R^2}}) = G M r/r_0^2
, \ee or \be \label{epseq} \epsilon^\prime = \frac {G M (r^3/r_0^3 - 1)} {r (1 -
1/\sqrt{1 - r_0^2/R^2}) (r^2/r_0^2 - 1)^{3/2}}. \ee

Our interest is not really in the function $\tt(r)$ because we
know that the proper time approaches infinity as the size of the
bubble approaches infinity, so $\tau \rightarrow \infty$ as $r
\rightarrow \infty$. We want to know where on the Penrose diagram
the brane ends up, or equivalently what light ray it approaches
asymptotically. For this purpose we introduce the coordinate $x^+$
defined by \bea
dx^+ = dt - \frac {dr}{1 - r^2/R^2}  \label{xplus}\\
x^+ = 0 \ \ {\rm at} \ \  t = r = 0. \eea Recall that

\be d\tau^2 = (1 - r^2/R^2)dt^2 - \frac {dr^2}{1 - r^2/R^2}, \ee
 so
\be dt^2 = dr^2 \left( \frac 1 {(1 - r^2/R^2)^2 } + \frac {\tau^{\prime
2}} {1-r^2/R^2} \right) \ee

The unperturbed solution asymptotically approaches the light ray
$x^+ = 0$; the perturbed solution is \be \Delta x^+ =
\int_{r_0}^\infty \frac {dr} {1 - r^2/R^2} (\sqrt{1 +
\tt^{\prime 2} (1 - r^2/R^2)} - 1). \ee Since 
$x^+ (\tau = \infty) = 0$
for the unperturbed solution, we
expand $\tt$ as above and keep only the first order piece.
 Thus \be x^+ (\tau = \infty) =
\int_{r_0}^\infty dr \frac {\epsilon^\prime} {\sqrt{1 - r^2/R^2 +
1/\tau^{\prime 2}}}. \ee Using (\ref{doteq}) and (\ref{epseq}),
this becomes \be x^+ (\tau = \infty) = \frac {G M r_0} {\sqrt{1 -
r_0^2/R^2} - 1} \int_{r_0}^\infty dr \frac {r^3/r_0^3 -1} {r^2 (r^2/r_0^2 -
1)^{3/2}} \ee The integral is clearly convergent, and up to a constant
can be done by dimensional analysis. The final answer is \be x^+
(\tau = \infty) =  - (const) \frac {G M} {1 - \sqrt{1 -
r_0^2/R^2}}. \ee

Looking back at the definition of $x^+$, equation (\ref{xplus}),
we see that negative $x^+$ means that now the brane is asymptotic
to a light ray which has positive $r$ at $t=0$. The brane has
moved out. The conformal diagram is shown in the main text,
figure \ref{massInside}.

\setcounter{equation}{0}
\section{de Sitter poles in the S-matrix}
We are in an unusual situation where we understand the
semiclassical solutions, but we don't know how to pick out a time
variable and define an energy. This is a problem because we want to
look for poles in the S-matrix as a function of energy. 
We do know how to compute the
action for semiclassical configurations. Our strategy will be to
compute the action semiclassically, and then get at the energy
through the back door by using the semiclassical formula
${\partial S \over \partial t} = -E$.

We are also able to extract information about S-matrix elements in
this way. In a one-dimensional quantum mechanics problem, the
S-matrix consists merely of phases. It is precisely true that the
phase shift, which is naturally thought of as a function of
energy, is the Fourier transform of the action, which is naturally
thought of as a function of the time: 
\be e^{i \delta(E)} = \int dt e^{i E t} e^{i S(t)}. 
\label{fourierET} \ee 
Here $\delta(E)$ is
the phase shift, and $S(t)$ is the action, as will be defined more
carefully below. The meaning of the formula is that the S-matrix
can be thought of in the usual energy representation or in the
time representation, and as always they are related by Fourier
transformation. In the energy representation, one computes energy
eigenfunctions and extracts the phase relationship between the
leftmoving and rightmoving waves. There is a meaningless constant
in deciding the zero of the phase shift. In the time
representation, one computes the amplitude to start from an
arbitrarily chosen point far to the right and come back to that
point in a time $t$. The amplitude is given by an integral over
paths as usual in quantum mechanics. We call this amplitude $e^{i
S(t)} $.  Again, there will be a
meaningless constant which depends on the choice of point. With
these definitions, equation (\ref{fourierET}) is exact, but in
practice we will estimate $S(t)$ using semiclassical techniques.

Semiclassically, we do the Fourier transform by saddle points. The
saddle point condition is \be
 {\partial S \over \partial t} = -E,
\ee which is a standard equation in classical mechanics.
Evaluating the function gives \be e^{i \delta(E)} =  e^{i (E t +
S(t))}, \ee or \be \delta(E) =  E t + S(t) = -t {\partial S \over
\partial t} + S(t) , \label{legendre} \ee which identifies
$\delta(E)$ and $S(t)$ as Legendre transforms of each other. This
is the standard relationship: the Legendre transform is the
semiclassical approximation to the Fourier transform.
The end result is that equation (\ref{legendre}) enables us to
turn semiclassical computations of the action into semiclassical
computations of the S-matrix.

\begin{figure}[!htb]
\center
\includegraphics [scale=.6, clip]
{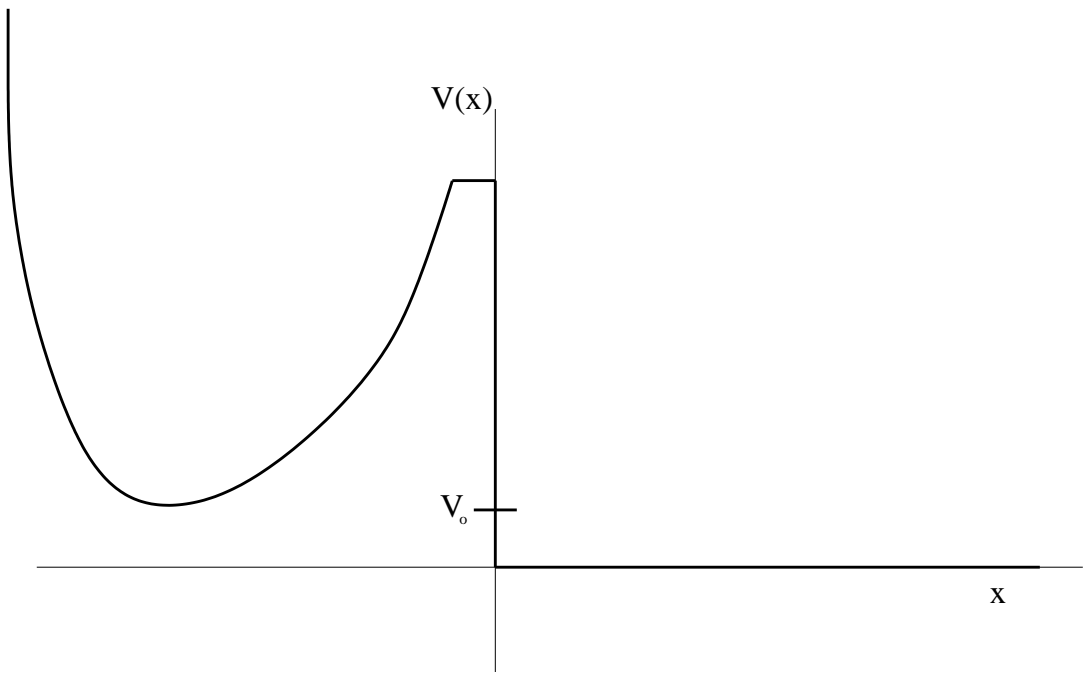} \caption{The S-matrix of a quantum mechanical particle moving
  in this potential exhibits a pole at the energy of the bound state.} 
\label{qmpot}
\end{figure}

To illustrate the technique, we begin with a simple quantum
mechanics problem. Consider a particle moving in the potential
shown in figure \ref{qmpot}.
We will approximate $S(t)$ by semiclassical methods and show that
our resulting S-matrix has the right behavior. For short times,
the only important trajectories are those which bounce off the
wall. The action is simply \be S(t) =\int dt \12 m \dot x^2 = \12
m \dot x^2 t = \12 m \frac {d^2} t. \ee Here we've chosen to start
and end our paths at $x = d$.

For long times, the only important trajectories are those which
tunnel into the well and stay there for a long time. The action
for these trajectories can be estimated as a sum of three
contributions: the action outside the well, the action inside the
well, and the action associated with tunneling through the
barrier. We will calculate each of these in a crude way. We could
do better for this problem, but the point is that crude estimates
provide information about the S-matrix.

The action associated with tunneling is an imaginary factor
$S_{tunneling} = i \gamma$. While the particle is inside the
well, it is described by a harmonic oscillator for which the
minimum potential energy is nonzero: \be S_{H.O.} = \int dt (\12 m
\dot x^2 - \12 k x^2 - V_0). \ee Over one period,
\be
\langle \12 m \dot x^2 \rangle = \langle \12 k x^2 \rangle
\ee
 so the only contribution is
from $V_0$. If the particle is in the well for many oscillations,
then the dominant contribution to the action is simply \be
S_{H.O.} = -V_0 t. \ee There will also be an oscillatory piece
which we choose not to compute.
 Finally, for these paths the
particle spends almost all of its time inside the well, so we
simply ignore the action due to the time it spends outside the
well. So we arrive at the formula for long times \be S(t) = -V_0 t
+ 2 i \gamma. \label{sformula} \ee Now let's see whether we get a
pole in the S-matrix. We Fourier transform to get \be e^{i
\delta(E)} = \int_0^\infty dt e^{i (E t - V_0 t + 2 i \gamma)}
 = {i e^{-2 \gamma} \over E - V_0},
\ee which has a pole at the resonance! In order to get the
imaginary part of the pole, we would have to do a little better,
but this example makes it clear that a crude semiclassical
understanding of the action can yield important information about
the S-matrix. If we adjust our energy
 scale so that $V_0 =0$, then $S(t)$ does not depend on $t$ for long times,
 as one can see from (\ref{sformula}). This corresponds to a pole at $E = 0$.

\begin{figure}[!!htb]
\centering \subfigure[]{\includegraphics
[scale=.6]{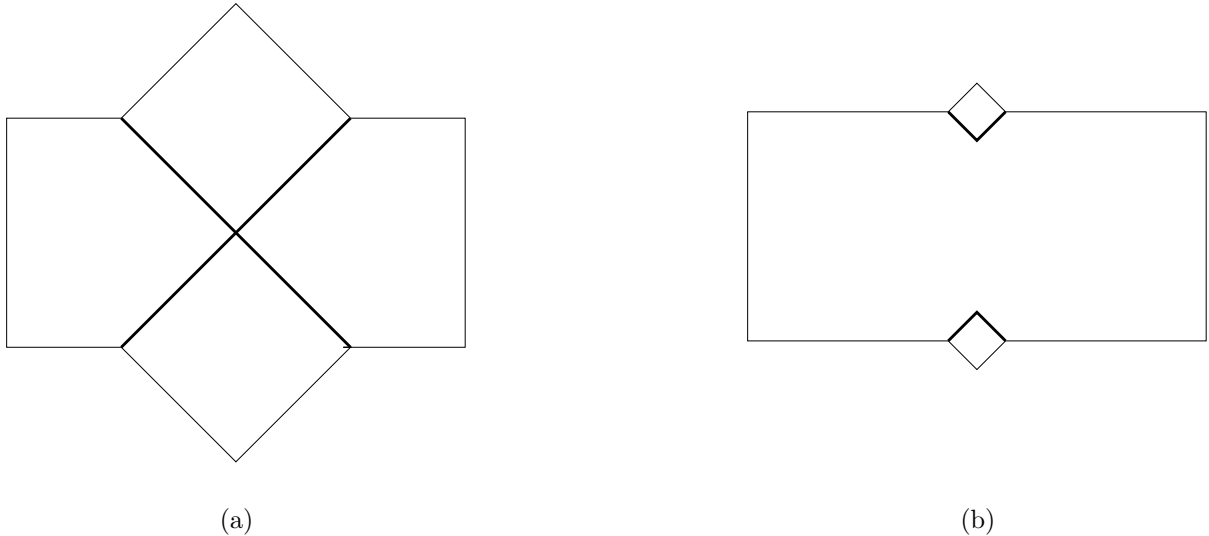} \label{zerotensionbubb} } \hfill
\subfigure[]
{\includegraphics [scale=.6]{zerotensiontunn}
\label{zerotensiontunn} } 
\caption{In the limit of zero brane tension, the domain wall always
  moves at the speed of light. Conformal diagrams for the classical solution (left) and the tunneling ``solution'' (right) are shown. }
\end{figure}

Now for the real problem. For simplicity, we take the
limit where the tension of the brane is extremely small, so that
it is essentially always moving at the speed of light. However,
our results will be general. In the zero tension limit, the
conformal diagram for the bounce solution looks like figure
\ref{zerotensionbubb}.  A tunneling solution is shown in figure
\ref{zerotensiontunn}.

\begin{figure}[!!!!htb]
\centering \subfigure[]{\includegraphics
[scale=.3]{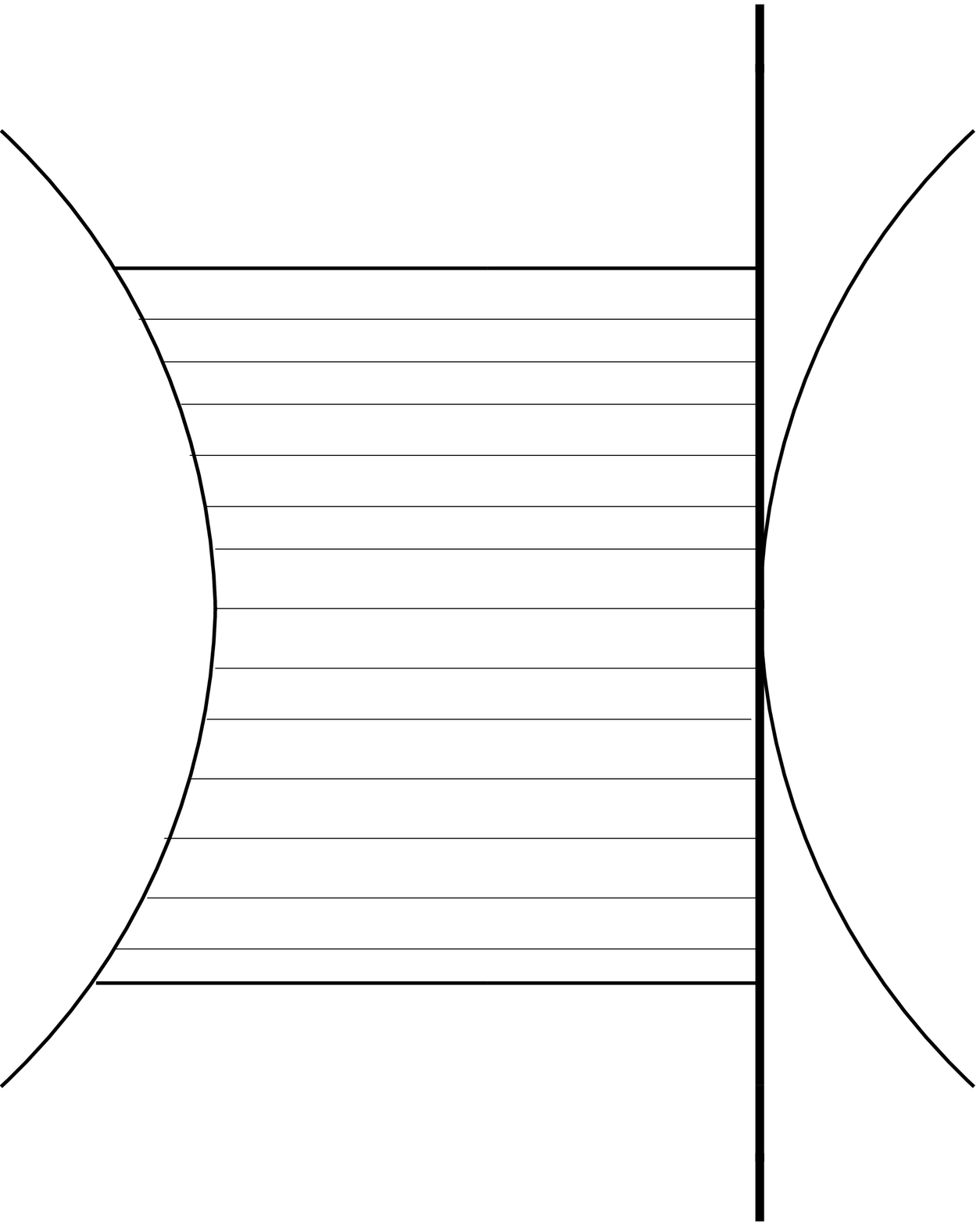} \label{gooduntunn} } \hfill
\subfigure[]
{\includegraphics [scale=.3]{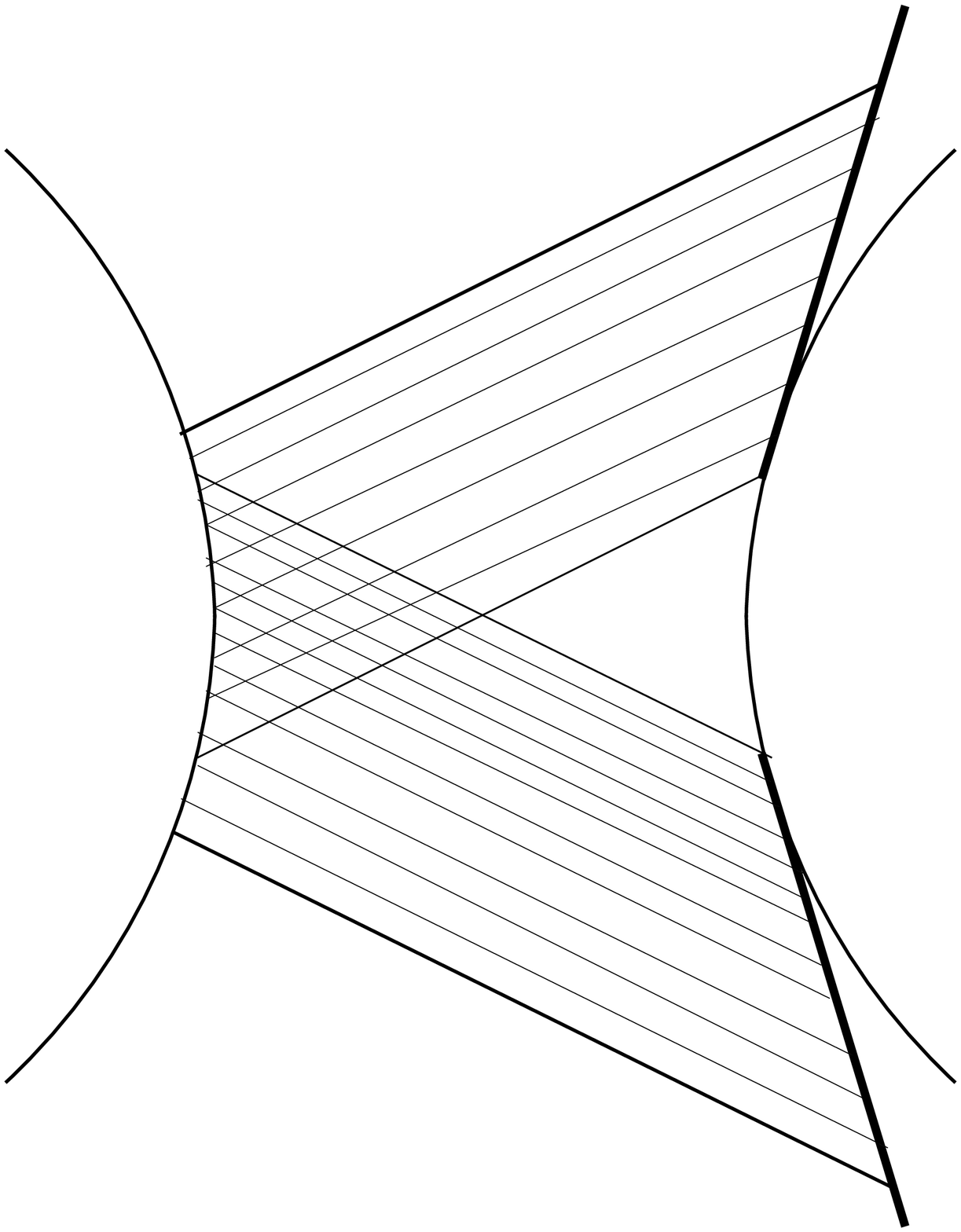}
\label{good} }
 \caption{At left, a slice through the embedding shows the classical solution in the limit of zero brane tension. The hyperboloid is sliced through by a plane (thick vertical line). At right, a tunneling solution. The hyperboloid  is sliced by two semi-infinite planes (thick rays). In both cases, time slices are shown. They are horizontal lines on the left; on the right they are boosted so they are not horizontal anymore. One can see that the area between the top and bottom horizontal lines on the left is the same as the area between the top and bottom boosted horizontal lines on the right, so the action is equal for the two configurations aside from the tunneling factor.}
\label{goody}
\end{figure}

The corresponding embedding is the hyperboloid cut by the plane
$x = 1$, as shown in figure \ref{gooduntunn}. We know the gravitational action is simply 
\be
S = {1 \over G} \int d^4x \sqrt g (R+ \Lambda).
\ee
 Since there is no timelike
Killing vector, it is not obvious which time variable to use in
defining $S(t)$. We will simply using  the embedding time since it
is easy to work with. In addition, we are interested in
an observer in the flat part of the bubble and the embedding time {\it is}
a Killing vector in the flat part of the geometry, so we might
expect that at very late times when the branes have gone most of
the way to infinity this time is the right one to use.

The action is infinite, because the space at the top and bottom of
the diagram becomes infinitely big. The action all comes from the
de Sitter part of the geometry, and is proportional to
\be
S \sim {\Lambda \over G} \int d^4x \sqrt g .
\ee
We choose to cut it off when the branes get out to a
distance $t_0$, which is also the time from the origin.

Now consider a tunneling solution. In the limit of zero brane
tension, the bubbles will nucleate at zero size. Consider for a
moment the two points from which the bubbles nucleate. There are
enough symmetries in de Sitter space that only the invariant
distance between the two points matters, so I can place them time
symmetrically and along the same spatial axis. Each individual bubble
nucleation is a boost of one which starts at t = 0. They each are
defined by planes which are tangent to the hyperboloid at the time
of bubble nucleation. A slice through the geometry is shown in figure
\ref{good}. It is natural to cut off the integral along a time slice
in the rest frame of the bubble. Then a geometric argument (figure
\ref{goody}) shows that
the action for the tunneling solution is exactly the same as the
action for the classical solution aside from the tunneling factor.

  The time delay can be arbitrarily long for tunneling solutions, so for
long times the action is NOT a function of the time delay. This is
exactly what happened in the quantum mechanics example when we chose
$V_0 = 0$. Just like in that example, the fact that the action is
independent of the time delay indicates a pole at $E = 0$.
 Since the action was not a function of the time delay, our confusion
 about which time to use was unimportant. A miracle has occured:
 We have been able to conclude not
only that all of these solutions have the same energy, but also
that they all have zero energy without having to precisely define
what we mean by time.

\end{appendix}

\end{document}